\documentclass[graphicx,11pt]{aastex}

\lefthead{}
\righthead{}

\begin{document}

\title{Kuiper belt: formation and evolution} 

\author{\textbf{A. Morbidelli$^{(1)}$, D. Nesvorn\'y$^{(2)}$}\\  
(1) Laboratoire Lagrange, UMR7293, Universit\'e de Nice Sophia-Antipolis,
  CNRS, Observatoire de la C\^ote d'Azur. Boulevard de l'Observatoire,
  06304 Nice Cedex 4, France. (Email: morby@oca.eu / Fax:
  +33-4-92003118) \\
(2) Southwest Research Institute,
  1050 Walnut St., Suite 300, Boulder, CO 80302, USA.\\
} 

\begin{abstract}

This chapter reviews accretion models for Kuiper belt objects (KBOs), discussing in particular the compatibility of the observed properties of the KBO population with the streaming instability paradigm.  Then it discusses how the dynamical structure of the KBO population, including the formation of its 5 sub-components (cold, hot, resonant, scattered and fossilized), can be quantitatively understood in the framewok of the giant planet instability. We also establish the connections between the KBO population and the Trojans of Jupiter and Neptune, the irregular satellites of all giant planets, the Oort cloud and the D-type main belt asteroids. Finally, we discuss the collisional evolution of the KBO population, arguing that the current size-frequency distribution below 100~km in size has been achieved as a collisional equilibrium in a few tens of My inside the original massive trans-Neptunain disk, { possibly with the exception of the cold population sub-component}.

\end{abstract}

\section{Introduction}

The Kuiper belt provides a very rich set of clues to unveil how the Solar System formed and evolved. The very existence of objects so far from the Sun is significant. The outer edge of the classical Kuiper belt, together with the extremely low mass of the cold population, suggests that the Kuiper belt represents a relic of planetesimal formation at the maximal distance where such formation process worked. The size distribution of the Kuiper belt objects (KBOs), their colors, density and binarity provide important constraints on this process. Section~\ref{accretion} will detail on this. 

The complex orbital structure of the Kuiper belt, with at least 5 distinct components partially overlapping (cold classicals, hot classicals, resonant objects, scattered disk and fossilized scattered disk - see Gladman et al. 2008 and Chapter I for formal definitions) shows that a great deal of dynamical sculpting occurred, that profoundly transformed the pristine disk of planetesimals. Reconstructing the origin of the Kuiper belt orbital structure provides key information on the complex dynamical evolution of the giant planet system, as described in Sect.~\ref{structure}. Understanding this evolution also provides predictions on the existence of other KBOs outside of the Kuiper belt, such as the Oort cloud objects, Trojans and irregular satellites of the giant planets, even some asteroids, as reviewed in Sect.~\ref{other}.  

Finally, the orbital evolution is coupled with the collisional evolution. The existence of delicate features, like wide binaries, provides constraints on how intense the collisional evolution was. This, in turn, has implication for the size below which KBOs are more likely to be collisional fragments rather than primordial planetesimals. This will be discussed in Sect.~\ref{collisions}

Since the time of the first KBO book in 2008 our knowledge of the Kuiper belt and our understanding of its formation and evolution have significantly evolved. The five dynamical classes had already been identified back in 2008, but their quantitative characterization has been improved since, allowing for more constraining tests of the models of the belt's primordial sculpting (Petit et al., 2011; Lawler et al., 2018a). 

The new models of planetesimal formation based on the streaming instability (Youdin and Goodman, 2005; Johansen et al. 2007) had just started to emerge, but their relevance for the Kuiper belt had not yet been understood. It was nevertheless noted that planetesimal formation by gravitational instability -proposed in Goldreich and Ward (1973), dismissed in Weidenschilling (1995) and resurrected by Youdin and Shu (2002) with the concept of a critical solids/gas density ratio- could change completely the picture set by the collisional coagulation scenario. Now a lot more is known on these instability processes and comparisons with the Kuiper belt properties can be done, as discussed in Sect.~\ref{accretion}.  

The size distribution of the Kuiper belt was poorly known below $D\sim 100$ km; the existence of a knee was still debated (Petit et al., 2008). Thus, it was difficult to assess whether the KBO population shows evidence for a collisional equilibrium, like the asteroid belt (Kenyon et al., 2008).  Today deep surveys (Fraser et al., 2014) and crater counts on Pluto and Charon provide information about the size distribution of KBOs down to about a couple of kilometers in diameter (Robbins et al., 2017), so the possibility of a collisional equilibrium can be tested (Sect.~\ref{collisions}). 

The notion that a phase of orbital instability of the giant planets (a.k.a. the Nice model -Tsiganis et al., 2005; Gomes et al., 2005; Morbidelli et al., 2007) should have occurred and played a major role in sculpting the Kuiper belt was already well established (Morbidelli et al., 2008), but the reproduction of the main structures of the Kuiper belt was still sketchy (Levison et al., 2008). Moreover, the possible evolutions of the giant planets during the instability were much less constrained than they are today (Nesvorn\'y and Morbidelli, 2012). 

With the model of a giant planet instability, it was already understood that some KBOs should survive today very far from their parent reservoir, as Trojans of Jupiter (Morbidelli et al., 2005) or irregular satellites of the giant planets (Nesvorn\'y et al., 2007). These capture models are now much more precisely tested. Moreover it has become clear that the Oort cloud and primitive D- or P-type asteroids are also related to the KBO population (Sect.~\ref{other}). 

This chapter will provide a review of our current understanding of the origin and evolution of the Kuiper belt, focusing on the advances achieved in the last decade.

\section{Accretion of KBOs}
\label{accretion}

In 2008, it was generally believed that KBOs (and large asteroids as well) formed by progressive collisional coagulation of primordial sub-km planetesimals. The origin of these small planetesimals was unexplained. {It is in fact known that dust coagulation is effective in building particles up to a few cm in size, but this process is then blocked by a number of effective barriers (the so-called bouncing, fragmentation and drift barriers; see Weidenschilling, 1977; Brauer et al., 2008; G{\"u}ttler et al., 2009;  Birnstiel et al., 2016).}  The review by Kenyon et al. (2008) extensively described the statistical coagulation-fragmentation codes employed to simulate the growth of KBOs from sub-km planetesimals and their final size distribution. Relatively little work has been done on collisional coagulation since (e.g. Schlichting and Sari, 2011; Kenyon and Bromley, 2012; Schlichting et al., 2013; Shannon et al. 2016). 

\begin{figure}[t!]
\centerline{\includegraphics[height=7.cm]{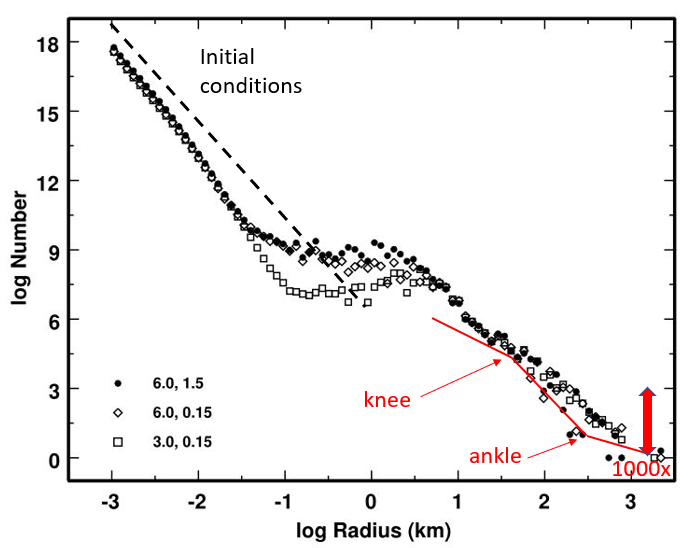}}
\caption{\small The initial (dashed line) and final (open and filled symbols, from three simulations assuming different strength parameters) size distributions in the Kuiper belt in the collisional coagulation model of Kenyon and Bromley (2004) -taken from their Fig. 9. The red lines sketch the currently observed size distribution. The vertical arrow indicates that the primordial distribution needs to be scaled up from the current one by a factor $\sim 1,000$. Coagulation models, therefore, fail producing the original size distribution of the trans-Neptunian disk. }
\label{Kenyon}
\end{figure}

Although the results somewhat differ from one coagulation model to the other, depending for instance on the assumptions on the strength of planetesimals and on the evolution of Neptune, the general trends are well summarized in Fig.~\ref{Kenyon}, taken from Kenyon and Bromley (2004). In essence, a handful of Pluto-size objects are formed, from which follows a steep single-slope size distribution down to small sizes unobserved with current telescopes. The currently observed KBO size distribution is different and sketched by the red lines in the figure. It is not a single power-law: it has a ``knee'' at about $R_{knee}\sim 50$~km where it turns from shallow (for $R<R_{knee}$) to steep (for $R>R_{knee}$) and an ``ankle'' at $R_{ankle}\sim 300$--400~km  where it turns to being very shallow (for $R>R_{ankle}$). Obviously, this last feature is only visible in the hot Kuiper belt/scattered disk, because the cold belt does not contain objects with $R>R_{ankle}$. The model by Schlichting et al. (2013) works better than the one presented in the figure in that it features a turnover in the size distribution around $R_{knee}$. To do so, however, it predicts huge waves in the size distribution at around $\sim 10$~km in size, which are ruled out by the crater size distribution on Pluto and Charon, as it will be shown later ({see Sect.~\ref{collisions}, which will also discuss the size distribution at km-scale}). 

Nevertheless, the main issue with coagulation models is not related to the observed broken power-law. By reproducing roughly the current number of objects, these models fail to reproduce the {\it original} number of objects in the trans-Neptunian disk. As it will be discussed in Sect.~\ref{structure}, there is ample evidence that the original disk population was 100-1,000 times more massive than the current scattered (Duncan and Levison, 1997; Brasser and Morbidelli, 2013; Nesvorn\'y et al., 2017) and hot populations (Nesvorn\'y, 2015b), respectively and that the number of Pluto-size objects was 1,000-4,000 (Stern, 1991; Nesvorn\'y and Vokrouhlick\'y, 2016). In a nutshell, all collisional coagulation models fail to produce enough big objects by orders of magnitude.    

Given the difficulty of collisional coagulation models in producing a large number of big objects, {as well as the problem of the origin of the initial sub-km planetesimals assumed in these models that we mentioned at the beginning of this section}, a large body of work has been developed in the last 13 years on the possibility to produce large bodies directly from dust aggregates (dubbed ``pebbles'') via the so-called streaming instability.

Although originally discovered as a linear instability (Youding and Goodman, 2005; Jacquet et al., 2011), this instability in the non-linear regime generates even more powerful effects, which can be qualitatively explained as follows. The key factors are the speed difference between gas (in a slightly sub-Keplerian rotation) and solid particles and the back-reaction of solids onto gas. Thus, the differential speed causes gas-drag onto the particles and the friction exerted from the particles back onto the gas accelerates the gas and diminishes its difference from the Keplerian speed. Consequently, if there is a small over-density of particles, the local gas is in a less sub-Keplerian rotation than elsewhere; this in turn reduces the local headwind on the particles, which therefore drift more slowly towards the star. As a result, an isolated particle located farther away in the disc, feeling a stronger headwind and drifting faster towards the star, eventually joins this over-density region. This enhances the local density of particles and reduces further its radial drift. It is easy to see that this process drives a positive feedback, i.e. and instability, where the local density of particles increases exponentially with time. 

The particle clumps generated by the streaming instability, if dense enough, can become self-gravitating and contract to form planetesimals. Numerical simulations of the streaming instability process (Johansen et al., 2015; Simon et al., 2016) show that planetesimals of a variety of sizes can be produced, but those that carry most of the final total mass are those of $\sim$100~km in size. Thus, these models suggest that planetesimals form (at least preferentially) big, in stark contrast with the collisional coagulation model in which planetesimals would grow progressively from sub-km objects by pair wise collisions. The fact that the size of 100~km is  prominent (i.e. the ``knee'') in the observed size-frequency distributions of both asteroids and Kuiper-belt objects is a first important support for the streaming instability model.   

\begin{figure}[t!]
\centerline{\includegraphics[height=7.cm]{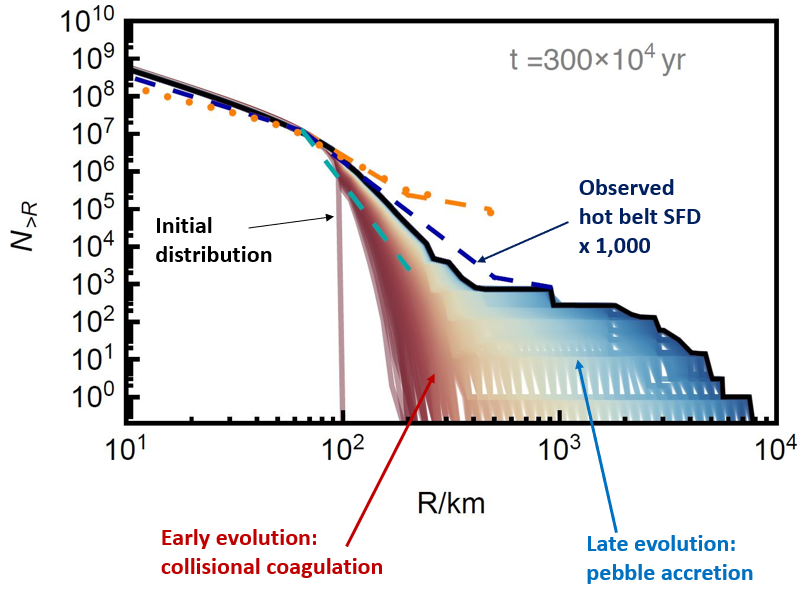}}
\caption{\small The evolution of the size distribution in the trans-Neptunian disk, centered at 25~AU, over 3~My, starting from the initial size distribution predicted in the streaming instability model of Johansen et al. (2015) and progressing via mutual collisions and pebble accretion. Here the initial mass of planetesimals was 25 Earth masses. Pebbles had a size of 0.4~mm. The disk's aspect ratio, {that is the thickness of the disk of gas relative to the stellar distance,} was 5\% and its turbulence was described by the $\alpha$-prescription of Shakura and Sunyaev(1973) with $\alpha=10^{-5}$. From Lambrechts and Morbidelli (2016).}
\label{Lambrechts}
\end{figure}

An attempt to reproduce the size distribution of the Kuiper belt starting from the size distribution produced by the streaming instability model has been done in Lambrechts and Morbidelli (2016) and is illustrated in Fig.~\ref{Lambrechts}. The planetesimals produced in the streaming instability (marked ``initial distribution'' in the figure) have initially a phase of collisional coagulation, which brings the largest objects to acquire a size of 300-400~km in radius. These objects then start to efficiently accrete individual pebbles, here assumed to be 0.4~mm in size, as the latter migrate radially in the disk. The acceleration in growth due to this process of {\it pebble accretion} (Lambrechts and Johansen, 2012, 2014) produces the  ``ankle'' and the terminal, shallow part of the observed size distribution. The total number of objects, about 1,000 times larger than at the present time for any size, is also successfully reproduced. Thus, for the first time we have a satisfactory model for the primordial size distribution in the trans-Neptunian disk, although it should be noticed that the values of the model parameters (pebble size and flux, disk turbulence and aspect ratio, mass of the initial planetesimals produced by the streaming instability) have been chosen to achieve a best fit. 

A strong support for the streaming instability model comes from Kuiper belt binaries. A large fraction of the cold population KBOs are binary. These binaries are typically made of objects of similar size and identical colors (see Noll et al. 2008 and Chapter 9 for reviews). Accounting for the fact that wide binaries can be dissociated by collisions (Petit and Mousis 2004; Nesvorn\'y et al., 2011) it is possible that {nearly} all initial cold population KBOs were binary (Fraser et al., 2017). {The discovery that Ultima Thule - a member of the cold population that looked like a single object from HST observations (Benecchi et al., 2019) - is in reality a contact binary strengthens this conclusions, suggesting that planetesimals are born twins, although with a range of possible separations.  The fact that the fraction of binary KBOs is much lower in the hot population does not contradict this interpretation: the deficiency of hot KBO binaries can be understood because these bodies should have undergone close encounters with Neptune in the past, dissociating most of the binaries (Parker and Kaavelars, 2010; see Sect.~\ref{hot}). The same is true for Centaurs, which are objects currently undergoing planetary encounters: only tight binaries can survive (e.g. the Typhon/Echidna binary; Noll et al., 2006). Comet-size objects, instead, even if they had formed as binaries, would have been dissociated or collapsed as bi-lobed objects by a combination of collisions and planetary encounters  (Nesvorn\'y et al., 2018b; see also Sect. 5). In summary, only the cold population can be diagnostic of an initial high binary fraction of planetesimals.}

 Nesvorn\'y et al. (2010) showed that the formation of a binary is the natural outcome of the gravitational collapse of the clump of pebbles formed in the streaming instability, if the angular momentum of the clump is large. Their simulations show that the typical semi-major axes, eccentricities and size ratios of the observed binaries are well reproduced by the model. The identity in colors between the two components (Benecchi et al. 2009) is a natural consequence of the fact that both components are made of the same material. This is a big strength of the model because such color identity cannot be explained in any capture or collisional scenario, given the observed intrinsic difference in colors between any random pair of KBOs (even within the cold population). 

Additional evidence for the formation of equal-size KBO binaries by streaming instability is provided by the spatial orientation of binary orbits. Grundy et al. (2019) have recently tabulated KBO binaries with known orbits. For about 20 of these binaries in the cold classical Kuiper belt the ambiguity between the true orbit and its mirror through the sky plane has been broken. They show a broad distribution of binary inclinations with $\simeq$80\% of prograde orbits ($i_{\rm b}<90^\circ$) and $\simeq$20\% of retrograde orbits ($i_{\rm b}>90^\circ$). To explain these observations, Nesvorn\'y et al. (2019) analyzed new high-resolution simulations and determined the angular momentum vector of the gravitationally bound clumps produced by the streaming instability. Because the orientation of the angular momentum vector is approximately conserved during collapse (Nesvorn\'y et al., 2010), the distribution obtained from these simulations can be compared with known binary inclinations. The comparison shows that the model and observed distributions are indistinguishable from each other (Nesvorn\'y et al., 2019). This clinches an argument in favor of the planetesimal formation by the streaming instability and binary formation by gravitational collapse. For comparison, Goldreich et al. (2002) proposed that binaries formed by capture during the coagulation growth of KBOs. Their L$^2$s mechanism predicts retrograde orbits with $i_{\rm b} \sim 180^\circ$ and their L$^3$ mechanism predicts a 3:2 preference for retrograde orbits (Schlichting and Sari, 2008). These capture models can therefore be ruled out.      

Davidsson et al. (2016) developed a hybrid model of KBO formation. In their model large KBOs are formed by the streaming instability as described above. Small, comet-size objects, however, are formed after gas removal, by collisional coagulation of the remaining, not-yet accreted, pebbles. The motivation for invoking this bimodal accretion process is that the chemical analysis of comet 67P/CG by the mission ESA/Rosetta revealed that the internal temperature of the comet should never have exceeded 70K. This suggests that comets formed late, otherwise they would have been heated significantly by radioactive decay of short-lived nuclei, like $^{26}$Al {(Prialnik et al., 2009)}. If comets had formed after gas removal, i.e. at least several millions of years after time-0, this would not have been the case. The problem is that, after gas removal, the velocity dispersion of pebbles, stirred by the presence of large KBOs, grows relatively fast in time. In Davidsson et al. model, the  average  random  velocity  in  the  primordial  disk (from 15 to 30 AU) is $\sim 40$m/s during  the  first 25 My. Pebbles are assumed to coagulate during mutual collisions, but at these speeds, this is far from obvious. Moreover, the model produces ``only'' $4\times 10^{10}$ comets with diameter $D>2$~km, which is somewhat anemic compared to expectations. In fact,  dynamical models need about $2\times 10^{11}$ objects with $D>2$~km originally in the trans-Neptunian disk to obtain a scattered disk and an Oort cloud sufficiently populated to explain the currently observed fluxes of short-period and long-period comets (Nesvorn\'y et al., 2017; Brasser and Morbidelli, 2013; Volk and Malhotra, 2008; Duncan and Levison, 1997). 

An implicit assumption in Davidsson et al. is that the streaming instability would form objects early (so vulnerable to radioactive decay). Hence, the requirement to form comets at a late time prompts to search for an alternative model of comet formation. But this assumption may not be correct. From the observational point of view, there are indications that also large bodies did not heat up substantially. Comet Hale-Bopp, one of the largest comets ever observed, with a diameter perhaps of 70~km, is also one of the richest in super-volatiles such as CO (Biver et al., 1996). Had this comet suffered substantial heating, these super-volatiles would have been rapidly lost. Moreover, many KBOs with $D<600$~km have bulk densities smaller than 1 g cm$^{-3}$ (Brown, 2013), suggesting an undifferentiated structure which, again, points to a low amount of internal heating, hence late formation. If big KBOs formed by the streaming instability, this suggests that the instability occurred relatively late in the lifetime of the disk, not early. 

Indeed, from the modeling side it has been shown that triggering the streaming instability requires a ratio between the surface densities of solids and gas which is 2-5 times the solar ratio (Johansen and Youdin, 2009; Yang et al. 2017), the exact value depending on parameters, such as the particle size. This enhanced ratio could be achieved locally by piling up radially-drifting particles in some specific ring of the disk (Drazkowska et al., 2016; Ida and Guillot, 2016; Schoonenberg and Ormel, 2017; Drazkowska and Alibert, 2017). Or it could be achieved globally in the disk at a much later time when a substantial fraction of the gas was removed by photoevaporation (Carrera et al., 2017). This predicts a dichotomy of formation ages of planetesimals. In the inner solar system, it is confirmed by the measurement of formation ages of meteorites, which show a clear dichotomy between iron meteorites, daughter products of planetesimals accreted in the first $\sim 10^{5}$y (Kleine et al., 2005) and chondrites, which formed 2-3 My later (Villeneuve et al. 2009). In the outer Solar System, there is evidence that planetesimals formed very early in Jupiter's region, given that this planet reached $\sim 20$ Earth masses within 1 My (Kruijer et al., 2017), which argues for the pile-up of solids at/near the snowline (Ida and Guillot, 2016; Schoonenberg and Ormel, 2017; Drazkowska and Alibert, 2017) while, as said above, KBOs formed late. If the streaming instability in the trans-Neptunian disk occurred late, then comets may have formed by the streaming instability as well, either directly (Blum et al., 2017) or indirectly, namely as collisional fragments of larger objects formed by the streaming instability (see Sect.~\ref{collisions}). In either case we think that there is no need of invoking a specific mechanism for the formation of comets as in Davidsson et al. (2016). 

{We cannot fail noticing, however, that there is a contradiction between the idea of a late formation of KBOs triggered by gas-removal and the results of Fig.~\ref{Lambrechts}. If it takes really 3~My of collisional evolution and pebble accretion to evolve the original size distribution to the one currently observed for large objects, as the figure suggests, it is unlikely that the streaming instability that formed the first KBOs was triggered by photo-evaporation of the gas. In fact, pebble accretion requires the presence of gas, so the gas had to remain for long time after the formation of the first KBOs. This important issue deserves further and more detailed investigation. {Perhaps the streaming instability produced directly bodies as large as 300-400~km in radius, so that a long phase of collisional evolution was not needed before that pebble accretion could operate.} 

An additional indication of how the streaming instability might have operated comes from the structure of the Kuiper belt. As it will be discussed in Sect.~\ref{structure}, the cold component of the Kuiper belt formed locally and never had a large mass (probably never exceeding a tenth of an Earth mass). That population ended within 47~AU with no planetesimal disk beyond this limit. Instead, inside of 30-35~AU the disk of planetesimals carried a considerable mass, perhaps 20-30 Earth masses. This is required to form a populated enough hot Kuiper belt and scattered disk, but also to drive the giant planets to their current orbits at the end of their dynamical instability.  All this suggests that the streaming instability never occurred beyond $\sim 45$~AU, probably because the threshold solid/gas density could never be achieved; between $\sim 30$ and 45~AU the streaming instability could occur, but only sporadically, forming a population of KBOs (the cold population) with a small total mass; within $\sim 30$~AU the streaming instability was efficient, converting pebbles into tens of Earth masses of planetesimals; and near the snowline the streaming instability was not just efficient, but could operate early on, despite of the large gas density, forming the precursors of the giant planet cores. This view can be qualitatively understood considering that pebbles drift radially due to gas drag (the outer disk is depleted in solid material and the inner disk in enriched, with a pile-up formed near the snowline - (Ida and Guillot, 2016; Schoonenberg and Ormel, 2017; Drazkowska and Alibert, 2017) and that the streaming instability is triggered when the solid/gas ratio exceeds a threshold value.

\section{Dynamical sculpting of the Kuiper belt}
\label{structure}

It was already clear in 2008 that the structure of the Kuiper belt is intimately related to the dynamical instability that the giant planets should have experienced after the removal of gas from the protoplanetary disk (Morbidelli et al., 2008). But, as we commented in the introduction, at the time the planets' instability had not yet been fully characterized and, consequently, the reproduction of the Kuiper belt structure was, at best, qualitative (Levison et al., 2008). Thus, below we start reviewing the major advances achieved since 2008 in our understanding of the planets' evolution and of their instability, then we focus on how the main structural features of the Kuiper belt can be understood quantitatively in that framework.

\subsection{Giant planet instability}
\label{Nice}

The idea of a giant planet instability has its roots in the work by Thommes et al. (2002) and was then formalized in the so-called Nice model (Tsiganis et al., 2005; Gomes et al., 2005). While Thommes et al. aimed simply at demonstrating that Uranus and Neptune could have formed closer to the Sun, the motivation of the Nice model was more profound and still valid today. The planets should have formed on coplanar and circular orbits. Their interaction with the gas and with the planetesimals should have contributed in damping their eccentricities and inclinations, if any. So, why are the planets on moderately eccentric and inclined orbits today? Only two mechanisms can excite the eccentricity and the inclination of the planets: resonance crossing (which, typically, affects only the eccentricities) and mutual close encounters. Hence the need for a phase of instability. 

In the original version of the Nice model published in 2005, the initial orbits of the planets were chosen ad-hoc, conveniently close to the 1:2 mean motion resonance between Jupiter and Saturn to trigger an instability during their planetesimal-driven migration (Tsiganis et al., 2005). But contemporary studies on planet migration in gas-dominated protoplanetary disks soon showed that these initial conditions were not realistic. Instead, because Jupiter should have migrated towards the Sun more slowly than Saturn and these two planets together should have migrated more slowly than Uranus and Neptune, the giant planets should have been captured in mutual mean motion resonances during the gas-disk phase and be in a multi-resonant chain when the gas was removed (Morbidelli et al., 2007).  Thus, the original Nice model was abandoned in favor of a new model with initial planetary orbits taken from the output of hydrodynamical simulations of planet migration (Morbidelli et al., 2007; Levison et al., 2011). Nevertheless, the original multi-resonant configuration is not univocally determined (for instance, it depends on the density of the disk of gas and its aspect ratio). 

A detailed exploration of the outcome of the giant planet instability depending on the initial configuration was done in Nesvorn\'y and Morbidelli (2012) (NM12 hereafter). 

\begin{figure}[t!]
\centerline{\includegraphics[height=5cm]{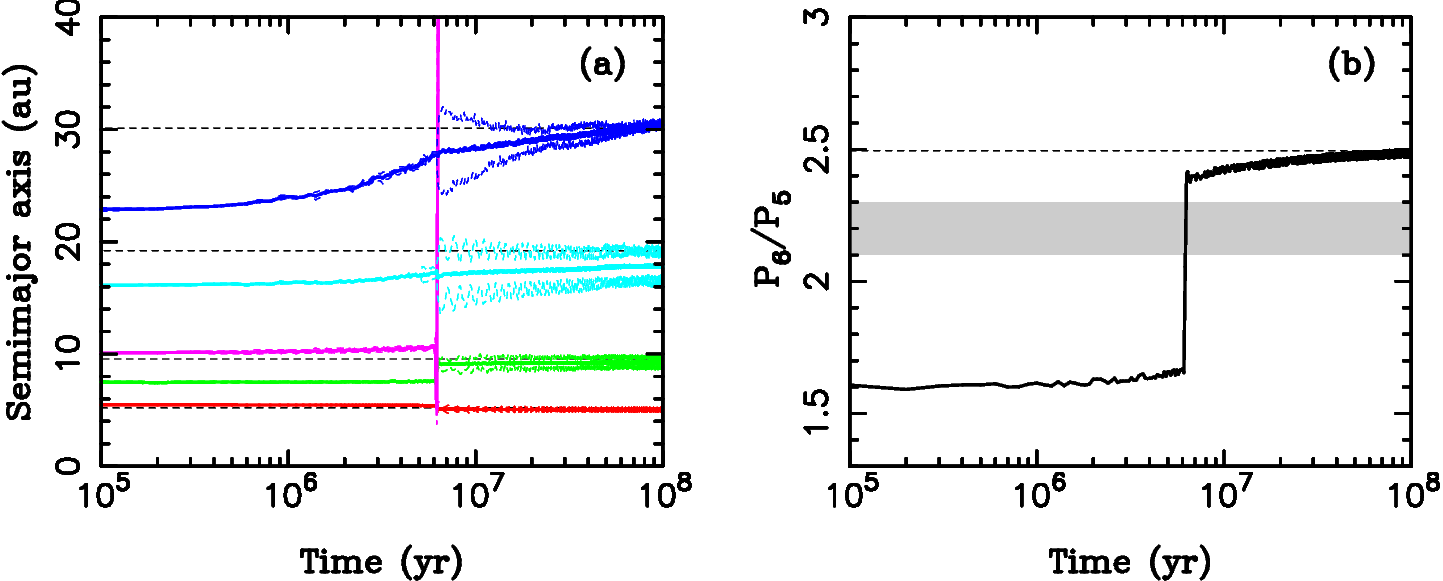}}
\caption{Orbital histories of the giant planets from NM12. Five planets were started in the 
(3:2,4:3,2:1,3:2) resonant chain, and $M_{\rm disk}=20$ Earth masses. Left panel: The semimajor axes (solid lines), 
and perihelion and aphelion distances (dashed lines) of each planet's orbit.  The horizontal dashed lines 
show the semimajor axes of planets in the present Solar System. The final orbits obtained in the model, 
including $e_{\rm 55}$, are a good match to those in the present Solar System. Right panel: The period ratio $P_6/P_5$. The dashed line shows $P_6/P_5=2.49$ corresponding to the present orbits of Jupiter and Saturn. The shaded area 
approximately denotes the zone where secular resonances with the terrestrial planets and asteroids 
occur. These resonances are not activated, because the period ratio `jumps' over the shaded area as Jupiter 
and Saturn scatter off of the ejected ice giant.}   
\label{Nf3}
\end{figure}

NM12 defined four criteria to measure the overall success of their simulations. First of
all, the final planetary system must have four giant planets (criterion A) with orbits that
resemble the present ones (criterion B), i.e. with final semimajor axes 
within 20\% to their present values, and the final eccentricities and inclinations not larger
than 0.11 and $2^\circ$, respectively. NM12 also required that the proper eccentricity of Jupiter $e_{55}$, {which is the most difficult to excite due to the planet's largest mass, is at least half of its current value, i.e. is larger than 0.022} (criterion C). Moreover, the period ratio between Saturn and Jupiter was required to evolve from $<2.1$ to $>2.3$ in $\ll 1$ My (Criterion D), because previous work had shown that this is necessary to preserve the inner asteroid belt (Morbidelli et al., 2010) and the terrestrial planet system (Brasser et al., 2009) if it already existed at the time of the giant planet instability. NM12 relaxed the condition that only the 4 known giant planets existed at the end of the gas-disk phase. They also tested configurations with 5 and 6 planets, with the sum of the masses of the rogue planets comparable to that of Neptune. We will refer below to Uranus, Neptune and the rogue planet(s) generically as ``ice giants''. All planets were initially in multi-resonant configurations {for the reasons explained above}, but different period ratios were tested. 

Figure~\ref{Nf3} shows an example of a successful simulation starting from a 5-planet configuration that satisfied all four criteria. The instability happened in this case about 6 My after the start of the simulation. Before the instability, the three ice giants slowly migrated by scattering planetesimals. The instability was triggered when the inner ice giant crossed an orbital resonance with Saturn and its eccentricity was pumped up. Following that, the ice giant had encounters with all other planets, and was ejected from the Solar System by Jupiter. Jupiter was scattered inward and Saturn outward during the encounters, with period ratio $P_{6}/P_{5} $ moving from $\sim 1.7$ to $\sim 2.4$  in less than $10^5$y (right panel of Figure \ref{Nf3}). The orbits of Uranus and Neptune became excited as well, with Neptune reaching an eccentricity of 0.15 just after the instability. The orbital
eccentricities were subsequently damped by interactions with the planetesimal disk (dynamical friction).
Uranus and Neptune, propelled by the planetesimal-driven migration, reached their current
orbits some 100 My after the instability. The final eccentricities of Jupiter and Saturn
were 0.031 (with a proper value $e_{55} = 0.030$) and 0.058 respectively. For comparison, the mean
eccentricities of the real planets today are 0.046 and 0.054.

\begin{figure}[t]
\centerline{\includegraphics[height=5cm]{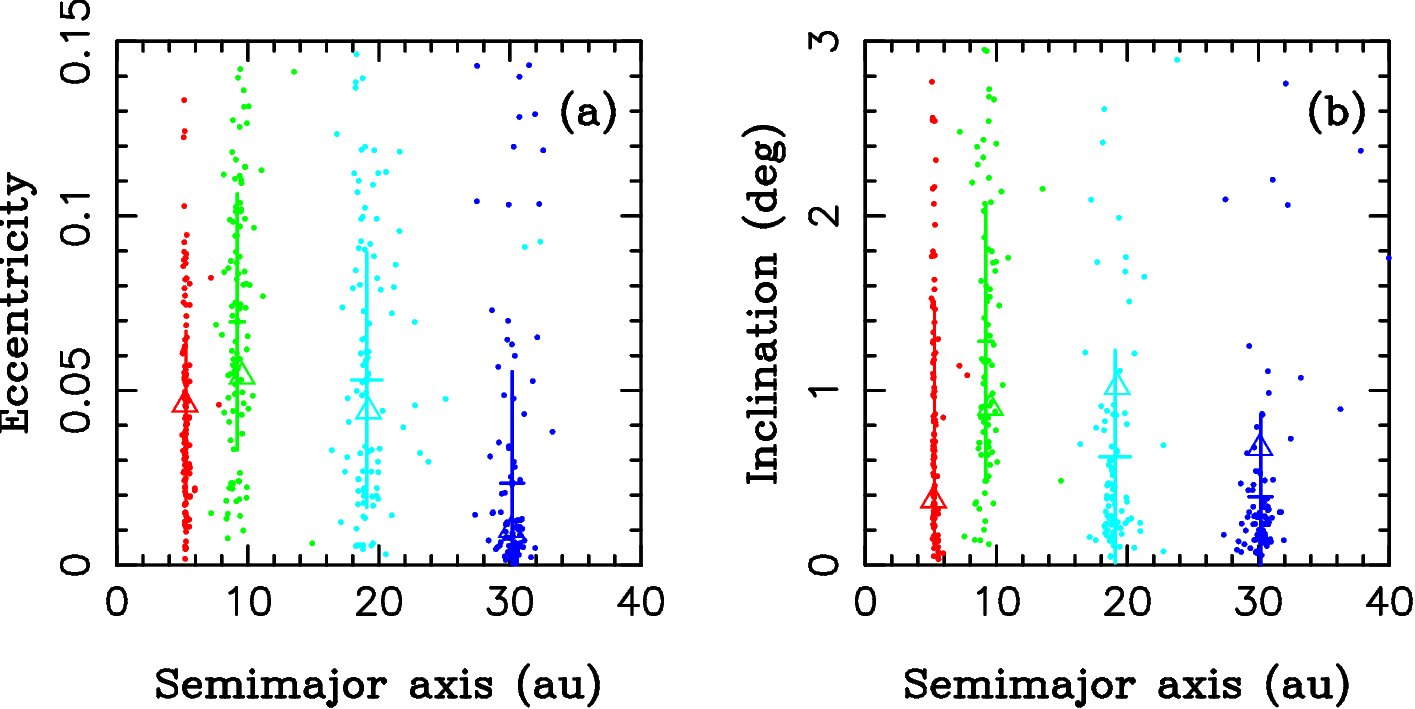}}
\caption{Final planetary orbits obtained in 500 simulations with five outer planets started in the (3:2,3:2,2:1,3:2) 
resonant chain and $M_{\rm disk}=20$ M$_\oplus$. The mean orbital elements were obtained by averaging the osculating 
orbital elements over the last 10 My of the simulation. Only the systems ending with four outer planets are 
plotted here (dots). The bars show the mean and standard deviation of the model distribution of orbital elements. 
The mean orbits of real planets are shown by triangles. Colors red, green, turquoise and blue correspond to Jupiter, 
Saturn, Uranus and Neptune.}   
\label{Nf4}
\end{figure}

NM12 found (see also Nesvorn\'y, 2011 and Batygin et al. 2012) that the
dynamical evolution is typically violent and leads to the ejection of at least one ice giant from the Solar System. Planet ejection
could be avoided if the mass of the planetesimal disk exceeds 50 Earth masses, but a
massive disk would lead to excessive dynamical damping and would drive the planets too far from each other. Thus, the dynamical simulations starting with
a resonant system of four giant planets have a very low success rate. In fact, NM12
did not find any case that would satisfy all four criteria in nearly 3000 simulations of
the four planet case. Consequently, either the Solar System followed an unusual evolution path
($<1/3000$ probability to satisfy criteria A-D), or some constraints are misunderstood, or there
were originally more than four planets in the outer Solar System.

Better results were obtained in NM12 when the Solar System was assumed to have five
giant planets initially and one ice giant was ejected into interstellar space by Jupiter (Figure~\ref{Nf3}). 
The best results were obtained when the ejected planet was placed into the external 3:2 or 4:3 resonance with Saturn and the mass of the planetesimal disk was 15-20 Earth masses. The range of possible outcomes is rather broad in this case
(Figure~\ref{Nf4}), indicating that the present Solar System is neither a typical nor expected result
for a given initial state, and occurs, in best cases, with a $\sim$5\% probability (as defined by
the NM12 success criteria). Notice that if it is assumed that each of the four NM12 criteria is satisfied
in 50\% of cases, and the success statistics are uncorrelated, the expected overall success rate is $0.5^4 = 0.063$, i.e.  6.3\%.

\subsection{Reproducing the main structures of the Kuiper Belt} 

Because in all versions of the Nice model the massive portion of the planetesimal disk had to have an edge around 30~AU, Levison et al. (2008) assumed that no planetesimals existed originally beyond this limit. The current Kuiper belt was empty. They showed that during the dynamical instability a fraction of the disk's planetesimal could be implanted into the Kuiper belt, and they proposed to identify the hot population with the bodies implanted from the inner part of the disk, more violently scattered by the giant planets, and the cold population with the bodies implanted from the outer part, more gently scattered. However, the predicted distribution of orbital inclinations of the hot population was found to be narrower than the one inferred from observations while the orbital excitation of the cold population was too high. The distinction between hot and cold populations was not sharp enough also in terms of initial location in the disk, and therefore it could not explain the  observed differences in spectral/color properties. Moreover, Parker and Kaavelars (2010) demonstrated that most KBO binaries would not have survived the encounters with Neptune required to transport the cold population to its current location. Finally, the resonant populations were overpopulated in Levison et al. (2008). Clearly, the Levison et al. paper was pioneer in attempting a global reconstruction of the Kuiper belt in the framework of the giant planet instability, but it needed to be improved.   Armed with a much better understanding of the evolutions of the giant planets that are compatible with the planets' current orbits (see sect.~\ref{Nice}), the origin of the complex structure of the Kuiper belt has been revised completely, as we review below for each of its 5 components.

\subsubsection{The hot population}
\label{hot}

The successful simulations of the Nice model like that presented in Fig.~\ref{Nf3} show that Neptune can move away from the other planets by planetesimal-driven migration by several AUs before the instability happens. Nesvorn\'y (2015b) realized that this pre-instability migration phase, which did not exist in the model considered by Levison et al. (2008), could be key for the appropriate excitation of the disk prior to its implantation in the hot Kuiper belt. Thus, in a series of simulations, he started Neptune at different positions between  20 and 30 AU and migrated it into
the disk on an e-folding timescale $\tau$ between 1 and 100 My, to test the dependence of the results
on the migration range and timescale. He found that reproducing the inclination distribution of the hot Kuiper belt requires that Neptune's migration was slow (with an e-folding timescale $\tau \ge 10$ My)
and long range (initial $a_{Neptune} < 25$ AU). The models with $\tau < 10$ My do not satisfy the inclination constraint, because there is not enough time for dynamical processes to raise inclinations. 

\begin{figure}[t]
\centerline{\includegraphics[height=5cm]{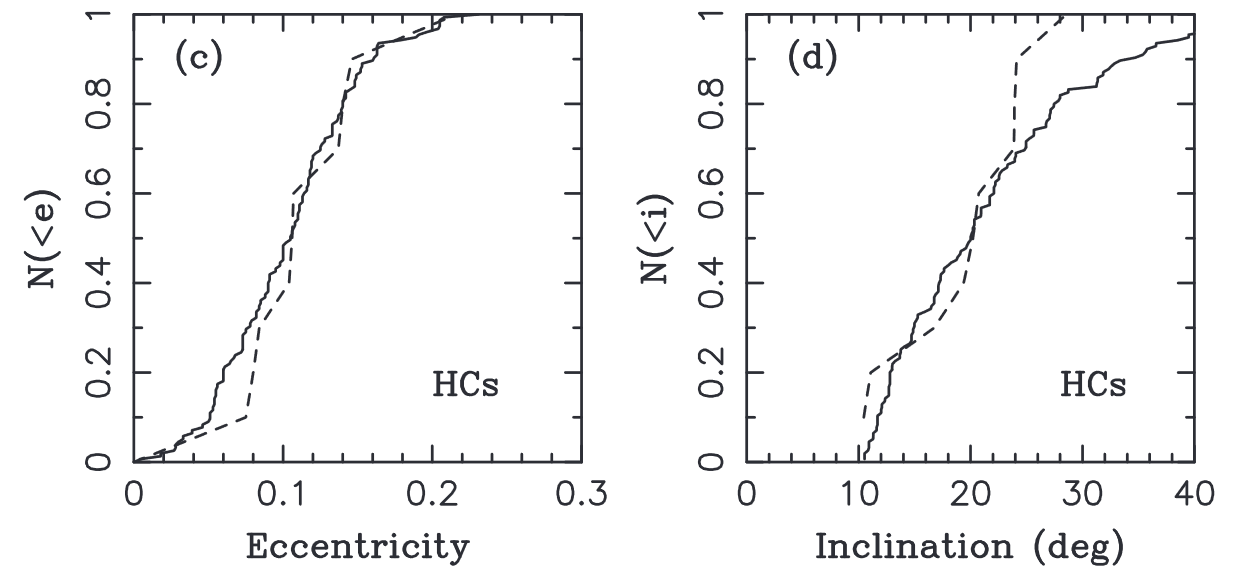}}
\caption{The cumulative distribution of eccentricities
 (left) and inclinations (right) for the hot Kuiper Belt.
The dashed lines show the actual CFEPS detections. The solid lines show the distributions of model bodies, detected by the CFEPS simulator. The comparison is done only for $i>10^\circ$  to avoid any potential contamination of the detection statistics from the cold population. From Nesvorn\'y (2015b).}   
\label{HKB}
\end{figure}

Fig.~\ref{HKB} shows the eccentricity and inclination distribution of the population implanted at $i>10^\circ$ in a simulation with Neptune starting at 24 AU and migrating with $\tau=30$~My. For a quantitative comparison with the observations {it is important to have an accurate estimate of the observational biases. For this reason, only the objects detected by the CFEPS survey have been considered because of the availability of an accurate survey simulator (Petit et al., 2011). By simulating the detection of synthetic objects generated according to the model, the observational bias is applied to the model distribution. The resulting distribution can then be directly compared with the observed one. The agreement between model and observations is remarkable, as shown in Fig.~\ref{HKB}. Unfortunately, it is not possible to extend the comparison to the objects detected by other surveys because of the lack of quantitative knowledge of the corresponding biases.}  

Only a small fraction ($\sim 10^{-3}$) of the disk planetesimals became implanted into the Kuiper belt in this kind of simulations. The implantation process is basically the one already described by Gomes (2003, 2011) and Gomes et al. (2005b). Namely, the objects are first scattered by Neptune, which pushes them outwards in semi-major axis and increases their inclination. Then, they are captured in some mean motion resonance. The complex secular dynamics in the resonance  decrease the objects' eccentricity and raise their perihelion distance, so that Neptune is no longer able to have close encounter with them. The continued migration of Neptune (and hence of its resonances) can finally drop these objects off resonance, so that they remain permanently trapped in the Kuiper belt region, preserving the inclination they previously acquired. 

The slow migration of Neptune required to reproduce the hot population represents an important clue about the original mass of the outer disk. For example, in the NM12 planetary migration/instability model where the trans-Neptunian disk extends from $\sim$23 to 30
au, $\tau \ge 10$ My implies $M_{disk} \sim 15-20$ Earth masses.

\subsubsection{The cold population}
\label{cold}

Parker and Kaavelars (2010) showed that the transport of the members of the cold Kuiper belt from smaller heliocentric distances via close encounters with Neptune, as proposed by Levison et al. (2008), would have dissociated most of the KBO binaries. A resonant transport during Neptune's migration, not involving close encounters with the planet had also been proposed (Levison and Morbidelli, 2003). But the point remains that the sharp differences in color (Trujillo and Brown, 2002) and size distributions (Fraser et al., 2010, 2014) between cold and hot KBOs suggest a different origin for the two populations. If the cold population had been transported from within 30 AU as advocated in Levison et al. (2008), given that this portion of the disk is also the source of the hot population (see Sect.~\ref{hot}), these differences would be difficult to understand. Thus, over the years, the idea that the cold population formed in situ became more and more popular in the Kuiper belt science community. 

Even if the cold belt formed in situ, the preservation of small eccentricities and inclinations is far from trivial, if Neptune experienced a phase of large eccentricity during the giant planet instability (Batygin et al., 2011; Ribeiro de Sousa al., 2018). But Neptune's eccentricity and inclination are never large in the NM12 models (i.e. $e_{Neptune} < 0.15$  and $i_{Neptune}<2^\circ$) so that there is no  excessive orbital excitation in the $>40$~AU region, where the cold population is expected to have formed.

On the other hand, simulations that don't over-excite the cold population also lead to a limited dynamical depletion, of only a factor of $\sim 2$ (Nesvorn\'y 2015a). This implies that the total mass of sizeable objects in the 42-47 AU region, currently of $3\times 10^{-4}$ Earth masses (Fraser et al., 2014),  was always small. It has been proposed that the cold population originally may have been significantly more massive and have lost most of its mass by collisional grinding (Pan and Sari, 2005), but the presence of loosely bound binaries places a strong constraint on how much mass can be removed by collisions (Petit and Mousis, 2004; Nesvorn\'y et al. 2011).
As we said in Sect.~\ref{accretion}, the existence of a population of large bodies with a total small mass may suggest that the streaming instability in this frontier region of the Solar System could occur only sporadically. 

\begin{figure}[t]
\centerline{\includegraphics[width=7cm]{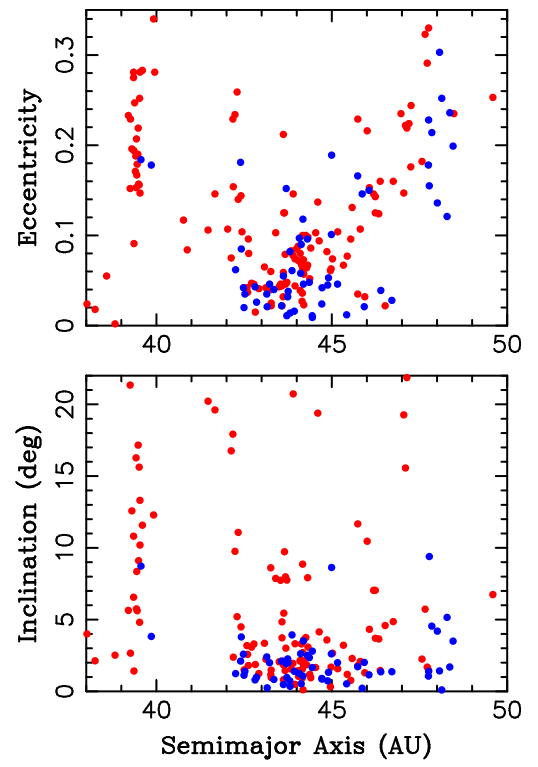}}
\caption{The distribution of eccentricities
 (top) and inclinations (bottom) vs. semimajor axis for the  Kuiper Belt objects detected by CFEPS (red) and those obtained in the model and detected by the CFEPS simulator. The apparent mismatch in inclination is due to the fact that here only the local population is modeled, so there is no hot population generated in the simulations. From Nesvorn\'y (2015a). }   
\label{CKB}
\end{figure}

A particularly puzzling feature of the cold population is the so-called kernel, a concentration of orbits with $a = 44$ AU, $e\sim  0.05$ and $i < 5^\circ$ (Petit et al. 2011). This feature can either be interpreted as a sharp edge beyond which the number density of the cold population drops (i.e. the original outer edge of the cold population), or
as a genuine concentration of bodies. If it is the latter, the kernel can be explained if Neptune's migration was interrupted by a discontinuous change of Neptune's semimajor axis
when Neptune reached $\sim 27.7$ AU (Petit et al. 2011, Nesvorn\'y 2015a), presumably due to the giant planet instability (see Fig.~\ref{Nf3}). Before the discontinuity happened, planetesimals located at $\sim 40$ AU were swept into the Neptune's 1:2 resonance, and were carried with the migrating resonance outward (Levison and Morbidelli 2003). The 1:2 resonance was at $\sim 44$ AU when Neptune reached $\sim 27.7$ AU. If Neptune's semimajor axis changed by a fraction of an AU at this point, perhaps because it had a close encounter with another planet, the 1:2 population would have
been released at $\sim 44$ AU, and would remain there to this day. The orbital distribution of
bodies produced in this model provides a good match to the orbital properties of the kernel
(Nesvorn\'y 2015a). 

Fig.~\ref{CKB} shows the resulting cold Kuiper belt in one of the simulations of Nesvorn\'y (2015a), featuring a Neptune's jump at 27.8 AU. The comparison with the observations of the CFEPS survey is done properly, by passing the model population through the CFEPS simulator. There are observed objects at inclinations larger than 5 degrees, that are absent in the model. {This is an additional indication that the hot Kuiper belt does not have a local origin, but rather comes from below 30 AU, as discussed in Sect.~\ref{hot}. Taking this consideration into account and restricting the comparison to $i<5$ degrees, the match between the model and the observations is excellent.} 

\subsubsection{The resonant populations}
\label{res}

The existence of resonant populations, particularly the group of Plutinos in the 2:3 resonance with Neptune, triggered the studies of the planetesimal-driven migration of this planet (Malhotra, 1993, 1995). These studies have been important from a perspective of history of science, because they have been the first to break the paradigm of in-situ formation of the planets (see also Fernandez and Ip, 1984). 

These works proposed that the resonant objects had been captured from a dynamically cold planetesimal disk swept by the resonances during Neptune's migration. But it was soon realized that the results of the model were at odds with observations. The resonant populations have an inclination distribution similar to that of the hot population, while they were expected to share the one of the cold population (Hahn and Malhotra, 2005). Their colors (Sheppard, 2012) and binary fraction (Noll et al., 2008) are also closer to the hot population properties. 

First Gomes (2003), then Levison et al. (2008), showed that objects can also be captured into resonance from the  population of planetesimals scattered by Neptune, and hence explained why resonant objects are more closely related to the hot population and the scattered disk than the cold population. Nesvorn\'y (2015b), while reproducing the inclination distribution of the hot population, also reproduced very accurately the inclination distribution of the Plutinos. These simulations showed that not only the 2:3 resonance is filled with objects, but also other resonances, such as the 1:2, 2:5, 1:3 etc. 

Models with smooth migration of Neptune invariably predict excessively large resonant
populations (e.g., Hahn and Malhotra 2005, Nesvorn\'y 2015b), while observations show that
the non-resonant orbits are in fact the most common (e.g., the classical belt population is $\sim $2-4
times larger than Plutinos in the 2:3 resonance; Gladman et al. 2012). This problem can
be resolved if Neptune's migration was grainy, as expected if
Neptune scattered massive planetesimals. The grainy migration acts to destabilize resonant
bodies with large libration amplitudes, a fraction of which ends up on stable non-resonant
orbits. Thus, the non-resonant/resonant ratio obtained with the grainy migration is up to $\sim 10$ times higher for the range of parameters investigated in Nesvorn\'y and
Vokrouhlick\'y (2016), than in a model with smooth migration. The best fit to observations
was obtained when it was assumed that the outer planetesimal disk below 30 AU contained
1000-4000 Pluto-mass objects. The combined mass of Pluto-class objects in the original disk was thus
$\sim 2$-8 Earth masses, which represents 10-50\% of the estimated planetesimal disk mass.

\subsubsection{The scattered disk}

The scattered disk is the Kuiper belt structure that was understood first (Duncan and Levison, 1997). Hence models of the scattered disk origin have not significantly evolved over the last decade. The scattered disk is also the least sensitive structure to the actual details of the past planets' dynamics. In fact, the scattered disk is what remains today of the population of objects scattered by Neptune since the beginning of the Solar System and which have not found a stable parking orbit. Duncan and Levison (1997) showed that this surviving population accounts for approximately 1\% of the original population scattered by the planet. This fraction may appear surprisingly high (for instance it is 10 times larger than the fraction captured on stable orbits in the hot population; see Sect.~\ref{hot}). This is due to the fact that many of the scattered disk objects are temporarily trapped in mean motion resonances with Neptune and therefore can live on non-encountering orbits for long time, before going back to scattered dynamics. 

More modern simulations, conducted in the framework of the Nice model (Brasser and Morbidelli, 2013; Nesvorn\'y et al., 2017) confirmed that the current scattered disk comprises 0.5-1\% of the original planetesimals in the trans-Neptunian disk. 

Therefore, this population is quite massive, compared to the other components of the Kuiper belt, as confirmed by observations (Trujillo et al., 2000). Possibly, only the fossilized scattered disk contains more objects (see below). But, because the scattered disk is intrinsically unstable and its population keeps decaying today, it dominates the flux of objects towards the giant planet region (Centaurs) and the inner Solar System (Jupiter family comets -JFCs). Therefore it can be considered as the reservoir of these populations. 

Several works have reproduced the orbital distribution of JFCs from the flux of objects from the scattered disk, provided that an appropriate physical lifetime is assumed for the comets (Duncan and Levison, 1997; Volk and Malhotra, 2008; Brasser and Morbidelli, 2013; Nesvorn\'y et al., 2017). These works concluded that the scattered disk should contain $\sim 2\times 10^9$ objects with $D>$2--3~km to explain the JFC population currently observed. Given the surviving fraction in the scattered disk, this implies that the original trans-Neptunian disk contained about $2\times 10^{11}$ of these objects (Nesvorn\'y et al., 2017).  

\subsubsection{The fossilized scattered disk}

The orbital structure of the Kuiper belt shows a population of objects that follows approximately the ($a,e$) orbital distribution of the scattered disk, but have larger perihelion distances, so that they are out of reach from Neptune's scattering action. Their orbital distribution suggests that the these bodies have been transported to large semimajor axis by encounters with Neptune. This means that they have been part of the scattered disk in the past but they are not part of it today.  Hence the name fossilized scattered disk. This name however is not unique. Gladman et al. (2008) used ``Detached Population'' instead. 

The observations suggest the existence of two sub-components of the fossilized scattered disk. One is present at all semimajor axes. It is basically the extension of the hot population to $a>50$~AU. The perihelion distances are typically within $\sim 45$~AU. Its origin is most likely related to (i) capture in mean motion resonance of scattered disk objects, (ii) increase in perihelion distance due to the secular dynamics in the resonance and (iii) drop-off resonance while Neptune (and its resonances) was still migrating (Gomes et al., 2005b; Gomes 2011). We have already described this process for the origin of the hot population. In fact, in this scenario, there is basically no difference between the hot population and this component of the fossilized scattered disk, apart from the arbitrarily chosen semimajor axis divide at 50~AU. Lawler et al. (2018b) concluded that, to have enough objects dropping off resonance, Neptune's migration should have been grainy and slow, in agreement to what was already concluded for the hot and resonant populations (see sect.~\ref{hot} and~\ref{res}). 

There seems to be a gap in perihelion distance between the two subgroups of the fossilized scattered disk (Trujillo and Sheppard, 2014).  In fact, the other sub-population of the fossilized scattered disk has perihelion distances $q>60$~AU. 
Three objects are currently known in this group: Sedna, 2012VP$_{113}$ and 2015TG$_{387}$. They are sometimes called ``Sednoids''. Given that these bodies are very big and can only be observed during a tiny arc of their orbital motion, they may represent the visible tip of the iceberg. In other words, the Sednoids may be the most massive component of the trans-Neptunian population.  

The fact that no bodies with perihelion distances comparable to those of the Sednoids  have been observed on orbits with semimajor axis smaller than 200~AU, despite the shorter orbital periods would make their discovery more probable, suggests that the population with very large $q$ exists only on wide orbits. This in turn suggests a specific trapping mechanism, operating only at large distances.  

This consideration immediately calls to mind the effect of passing stars and, more generally, the external potential due to the galactic environment of the Sun. In fact, the current galactic environment can raise the perihelion distance of objects, detaching them from the giant planets, if their orbits have $a\gtrsim 10,000$~AU (Dones et al., 2004). This is related to the formation of the Oort cloud, as we will see below (Sect.~\ref{Oort}). However, if the Sun formed in a stellar cluster, as most stars do, the external perturbations would have been stronger in the early times and they could have raised the perihelion distance of bodies already at a few hundred AUs in semimajor axis. Thus, since the discovery of Sedna, it has been proposed that Sednoids are objects ejected from the giant planet region when the Sun was still in its natal cluster (Morbidelli and Levison, 2004; Kenyon and Bromley, 2004b; Brasser et al., 2006, 2008, 2012). In alternative, it has been conjectured that these objects have been captured from the disk surrounding another star which had a close encounter with the Sun (Morbidelli and Levison, 2004; Kenyon and Bromley, 2004b; Jilkova et al., 2015). More recently, the planet IX hypothesis (Trujillo and Sheppard, 2014; Batygin and Brown, 2016) has opened the possibility that Sednoids have been trapped onto their current orbits by the perturbations exerted by this planet. However, the origin of the planet in first place would still be due to the same mechanisms: either ejection from the giant planet region and perturbations from the cluster or capture from another star. 

We end stressing that, if the formation of the Sednoids is due to the effects of a stellar cluster, the origin of these objects has to predate the giant planet instability. In fact, it is not possible to form the Sednoids and the Oort cloud at the same time. In a cluster such as the one required to capture the Sednoids, the Oort cloud would be unbound (Brasser et al., 2006, 2008). Because the Oort cloud exists, and the giant planet instability is the last event that scattered planetesimals around, it is necessary that the Sun was already in a galactic environment similar to the current one when the giant planets became unstable. Thus, presumably the ejection of the Sednoids occurred as the giant planets were still forming. Because gas was also present at that time and gas-drag prevents the ejection of comet-size bodies, this leads to the prediction that the Sednoid population is made only of big objects (Brasser et al., 2007). 

\section{Relationships with other populations of small bodies}
\label{other} 

The dispersal of the original trans-Neptunian planetesimal disk during the giant planet instability allowed some objects to be captured in regions of the Solar System quite far from the Kuiper Belt. These objects, strictly parented to those of the hot Kuiper belt, resonant populations and scattered disk comprise today the Trojan populations of Jupiter and Neptune, the irregular satellites populations of all four giant planets and the Oort cloud, but can also be found in the main asteroid belt. We focus on each of these sub-populations below.

\subsection{Trojan populations}
\label{trojans}

Jupiter and Neptune have Trojan bodies, i.e. small bodies leading or trailing the planet's position on its orbit. Saturn and Uranus don't have any Trojans but this is not surprising because their coorbital regions are unstable on long timescales (Nesvorn\'y and Dones, 2002). For long time, only the Trojans of Jupiter have been known; those of Neptune have started to be discovered only recently (Sheppard and Trujillo, 2010).  It was supposed that Trojans are a local population of planetesimals that got captured in co-orbital motion with the planet when the latter grew in mass (see Marzari et al., 2002). But this scenario cannot explain why the observed orbital inclinations of Jupiter's Trojans can reach very large values (now also true for Neptune's Trojans). 

Morbidelli et al. (2005) have been the first to show that the Trojans of Jupiter could have been captured during the instability of the giant planets. Their model explained the large orbital inclination of the Trojans, but was developed in the framework of the original version of the Nice model, then abandoned in 2007. As it was relying on a specific feature of the first Nice model, namely Saturn's crossing of the 1:2 resonance with Jupiter, it could not be applied to the new version of the model (Morbidelli et al, 2007; Nesvorn\'y and Morbidelli, 2012) and it could also not be applied to explain the origin of Neptune's Trojans. 

The capture of Trojans of both giant planets has been explained in the framework on the new Nice model by Nesvorn\'y et al. (2013) and  Gomes and Nesvorn\'y (2016), respectively. In essence, in both cases, when the considered planet had its last significant jump in semimajor axis (due to a close encounter with another planet) the planetesimals that happened to have semimajor axes similar to the post-encounter planet's semimajor axis got captured in the new co-orbital zone.

\begin{figure}[t]
\centerline{\includegraphics[height=7cm]{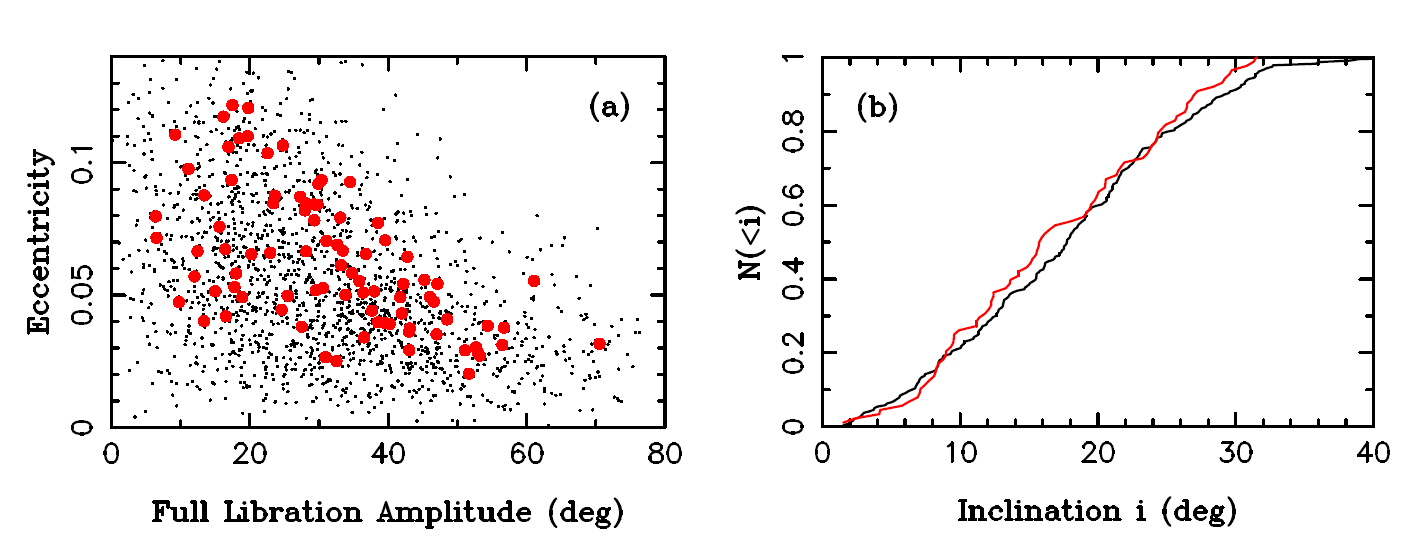}}
\caption{Left: The distribution of eccentricity vs. libration amplitude of observed Trojans (black dots) and planetesimals captured in the simulation (red). Right: comparison between the observed (black) and simulated (red) cumulative inclination distributions. From Nesvorn\'y et al. (2013).}   
\label{Trojans}
\end{figure}

In principle, the planet's orbital jump can produce Trojans from any source
reservoir that populated the planet's vicinity at the time of the jump. But, at the time of the giant planet instability, the overwhelmingly dominant source was without doubts the trans-Neptunian disk, undergoing dynamical dispersal. Using this source, the model reproduces in a quantitative manner the observed distribution of Jupiter's Trojans in terms of libration amplitude, eccentricity and inclination (Fig.~\ref{Trojans}). The model can also potentially explain the observed asymmetry in number of objects larger than a given size between the leading and trailing Jupiter's Trojan populations. In fact, the capture process is not symmetric: one of the two populations can be partially depleted, depending on the specific trajectory of the rogue planet during the close encounter with Jupiter that led to Trojans' capture. Simulations show asymmetries up to a factor of 50\%, sometimes in favor of one population, sometimes the other. The observed asymmetry is of $\sim 30$\% in favor of the leading population. 

Because the dispersed trans-Neptunian disk was also the source of the hot population (see sect.~\ref{hot}), this model leads to the strong prediction that Trojans and the hot population should have indistinguishable size and color distributions (Morbidelli et al., 2009). 

For the size distribution, this prediction was confirmed by the dedicated study of Fraser et al. (2014) for Jupiter's Trojans. The observed match can be hardly considered a coincidence, because the slope of the common distribution is very different from those of other populations, such as the asteroids or the cold Kuiper belt. 

For colors, however, it is observed that Jupiter's Trojans cover only part of the color distribution of the hot population. The reddest colors of the hot KBOs (those with B-R$\gtrsim$1.5) are missing among the Jovian Trojans. Wong and Brown (2016) explained this difference by assuming that the Trojans were resurfaced (for example, by sublimation of near-surface volatiles) as they moved from the trans-Neptunian disk (where surface temperatures are $\sim 50$ K) to Jupiter's orbit ($\sim 125$ K). But a recent work (Jewitt, 2018) shows that also Neptune's Trojans are missing the reddest objects. If this is confirmed when more Neptune's Trojans are discovered and characterized, it will invalidate the Wong and Brown's explanation, because the temperature of Neptune's Trojans is about the same as that of the KBOs. 

In this case, the explanation has to be dynamical. While the hot Kuiper belt population samples the entire disk dispersed by Neptune during its radial migration, the Trojans of Jupiter and Saturn just sample the dispersed population at the time of the last orbital jump suffered by these planets. If the original trans-Neptunian disk was characterized by a radial color gradient, with the grayer colors in the inner part and the redder ones in the outer part (Hainaut et al., 2012; Wong and Brown, 2016),  the absence of very red colors among the Trojans implies that the last orbital jumps of the planets occurred before that the very red component of the trans-Neptunian disk had started to be dispersed. { Numerical quantitative models are needed to test whether this is compatible with a jump of Neptune's orbit  when the planet was at $\sim 28$ AU, as discussed in sect.~\ref{cold}. One has to check whether the portion of the trans-Neptunian disk dispersed {\it after} this event could be a sufficient source of the red component of the hot population and whether these objects would have the correct inclination dispersion in the end. If the result were negative, the implication would be that Neptune was significantly closer to the Sun than 28~AU when its last orbital jump happened. In this case, the kernel of the cold Kuiper belt could not be related to Neptune's jump and most likely reveals the original outer edge of the cold population.} 

\subsection{Irregular satellites}

The irregular satellites of the giant planets are bodies that orbit the planet with inclined and eccentric trajectories, in sharp contrast with the co-planar and quasi-circular orbits of the regular satellite systems. 

The irregular satellite populations of the four giant planets are fairly similar to each other if the orbital semimajor axes are rescaled relative to the Hill radius of the planet (Jewitt and Sheppard, 2005). This led these authors to reject capture scenarios based on planet's growth or gas drag in the primordial planetary envelopes, because in this case the vastly different accretional histories of gas-giants (Jupiter and Saturn) and ice-giants (Uranus and Neptune) would imply very different irregular satellite populations. Instead, Jewitt and Sheppard favored unspecified mechanisms of conservative dynamics, because the capability of a planet to deflect an object scales with its Hill radius. 

This surprising intuition for a team of observers was confirmed by Nesvorn\'y et al. (2007), again in the framework of the Nice model. They showed that irregular satellites could be captured from the background planetesimal population generated by the dispersal of the trans-Neptunian disk during planetary encounters. More specifically, capture happened when the trajectory of a background planetesimal approached a pair of planets in mutual close encounter. In this three-body interaction the planetesimal may end up on a bound orbit around one of the planets, where it remains permanently trapped once the planets move away from each other. 

Modeling this mechanism in detail, Nesvorn\'y et al. (2014) found that
the orbital distribution of bodies captured during planetary encounters provides a good
match to the observed distribution of the irregular satellites around all four giant planets. 

Given the capture probabilities of irregular satellites in the numerical simulations and the expected size distribution in the original trans-Neptunian disk (deduced from the size distribution of the hot Kuiper belt and of the Trojans, multiplied by the inverse of their capture probabilities), the model predicts within a factor of 2 the size of the largest irregular satellite around each giant planet (Nesvorn\'y et al., 2007). The overall size distribution of irregular satellites, however, is much shallower than that of the hot Kuiper belt population. This discrepancy was explained by Bottke et al. (2010) by showing that the irregular satellite populations are strongly collisionally evolved. Due to the extremely high relative velocities between prograde and retrograde satellites, collisions lead to the super-catastrophic disruption of the targets. The fragments are therefore very small, so that the collisional evolution leads  to an equilibrium distribution with a slope of $\sim -2$ only for $D\lesssim 1$~km. Above this threshold, satellites are only destroyed, never regenerated, so that the resulting size distribution is extremely shallow (exponent of the cumulative size distribution larger than $-1$), as observed.

{Because the captures of irregular satellites and of Trojans are coeval, because they both occurring during the planetary close encounters,  this model predicts that the color distributions of these two populations of bodies should be indistinguishable. This prediction has been validated by Graykowski and Jewitt (2018), who compared the irregular satellite colors with the Jovian Trojan color distribution, finding no evidence for a significant difference.}

\subsection{The Oort cloud}
\label{Oort}

The main mechanism of formation of the Oort cloud was understood long ago (Shoemaker and Wolfe, 1984; Duncan et al., 1987) and was reviewed in a very detailed manner in Dones et al. (2004). In essence, planetesimals scattered by the giant planets onto orbits with semimajor axis $\gtrsim 10,000$~AU can have their perihelion distance increased enough to avoid a new close encounter with the planetary system during the subsequent revolutions. The perihelion-increase process is due to the tidal force exerted by the distribution of mass in the galaxy. Passages of nearby stars randomize the orbital distribution of the objects, giving to the Oort cloud its characteristic spherical structure. From time to time, the action of the galactic tide and of passing stars can decrease the perihelion distance of an object to $<5$~AU in less than an orbital period: in this case, a new long-period comet is born (Oort, 1950). 

Brasser et al. (2007) pointed out that gas drag prevents comet-size objects to be ejected onto orbits with semimajor axes typical of the Oort cloud. This is the case even if the comets are scattered off the disk's plane (where most of the gas is supposed to be) and if the disk itself is fairly small (with an outer edge of 50-100~AU). The implication of this result is very strong. Unlike what was previously supposed (Duncan et al., 1987; Dones et al., 2004) comets could not be sent to the Oort cloud during the formation phase of the giant planets, because the latter necessarily happened when there was still a significant amount of gas in the disk. Therefore, comets must have been sent to the Oort cloud only after gas removal. But, after gas removal, planetesimals were mostly beyond the giant planets and therefore they could only be scattered during the giant planet instability and the migration of Neptune, i.e. during the same phase that led to the formation of the hot Kuiper belt and of the scattered disk and to the capture of Trojans and irregular satellites. 

However, it is far from obvious that the dispersal of the trans-Neptunian disk during giant planet instability and migration could produce at the same time a scattered disk and an Oort cloud that are sufficiently populated to sustain the observed fluxes of Jupiter-family and long period comets. This has eventually been verified with numerical simulations in Brasser and Morbidelli (2013) and Nesvorn\'y et al. (2017). However, this result can be achieved only assuming that Oort cloud comets are intrinsically brighter (i.e. more active per unit surface) than JFCs, as independently advocated by Sosa and Fernandez (2011) by looking at the strength of non-gravitational accelerations. Using the Sosa and Fernandez (2011) conversion from comet brightness to nuclear size, the population estimates for the scattered disk and the Oort cloud are well reproduced in the Nice-model simulations, suggesting that both structures formed in the same event from the same parent population.

This conclusion leads to a strong prediction: although comets can be quite diverse from each other (because they sample a disk originally extended from $\sim 15$ to $\sim 30$~AU), statistically the populations of Jupiter family comets (coming from the scattered disk) and long period comets (coming from the Oort cloud) should be the same, given their common ultimate parent reservoir (the trans-Neptunian disk). {No systematic differences indeed appear in the $^{13}$C/$^{12}$C and $^{15}$N/$^{14}$N ratios between these two classes of comets (Jehin et al., 2008).}

However, for quite some time this prediction seemed to be at odds with the observed systematic difference in D/H ratio between long period comets and JFCs. But the ESA/Rosetta mission found that the JFC 67P-CG has a D/H ratio larger than any long period comet measured to date (Altwegg et al., 2015), suggesting that the previously claimed systematic difference was just apparent and due to small number statistics. Another apparent systematic difference between the two populations of comets is that carbon-depleted comets are only observed among JFCs (A'Hearn et al., 1995). This could be due to the facts that the observed carbon-bearing species are very volatile and that JFCs have suffered many more revolutions through the inner Solar System than long period comets, so it is not necessarily diagnostic of an original compositional difference between the two types of comets.

\subsection{Primitive asteroids}

Levison et al. (2009) pointed out that during the giant planet instability a fraction of the planetesimals coming from the trans-Neptunian disk should have been implanted in the main asteroid belt and in the 3:2 resonance with Jupiter (Hilda population). By analogy with the spectral properties of Jupiter's Trojans, classified as P- and D-type, Levison et al. proposed that all P- and D-type asteroids in the main belt and Hilda populations are dormant comets parented with the hot Kuiper belt. 

The implantation process has been revisited by Vokrouhlick\'y et al. (2016) in the framework of the most successful realizations of the Nice model of Nesvorn\'y and Morbidelli (2012). They found a  mean probability of $\sim 5 \times 10^{-6}$  for each trans-Neptunian disk body to be implanted into the asteroid belt. From the expected size distribution of the original trans-Neptunian disk, this capture probability leads to a number of large P-/D-type bodies in the belt ($D > 150$~km) consistent with observations. However it predicts a significant excess over the estimated population of smaller P-/D-types, because the size-distribution of the trans-Neptunian disk was steeper than that of the main asteroid belt.  This problem can be solved invoking the massive collisional destruction of small P-/D- type bodies, which is possible if they are more fragile than indigenous asteroids (Levison et al., 2009).

Unlike Levison et al. (2009), Vokrouhlick\'y et al. (2016) observed implantation of trans-Neptunain objects also in the inner asteroid belt ($a<2.5$~AU), where indeed some D-type asteroids are observed (DeMeo and Carry, 2014).

\section{Collisional evolution}
\label{collisions}

Given the dynamical history described in Sect.~\ref{structure}, it is clear that the KBOs suffered two stages of collisional evolution. The first stage was in the trans-Neptunian disk, very massive but with moderate dynamical excitation, and the second phase was during the dispersal of the disk and the implantation of some of its members in the various structures of the Kuiper belt. Given that the scattered is still decaying in population, there is no sharp transition between this second phase and today's environment, so both have to be treated together. 

Let's start our discussion from this second stage, because it is unavoidable that it happened and the associated dynamical evolution is now well characterized (see Sect.~\ref{structure}). The first stage is instead more uncertain, because the duration of the massive disk is poorly known. 

The collisional evolution during the dispersal of the trans-Neptunain disk till today has been modeled in Jutzi et al. (2017), from the dynamical simulations of Brasser and Morbidelli (2013). They showed that the effects of collisions is fairly moderate. The bodies with a catastrophic disruption probability of $\sim 1$ are those with $D\sim 4$ km. This means that bodies larger than 10~km (namely all visible KBOs) have been relatively unaffected. It is unlikely that the size distribution of the Kuiper belt populations has significantly evolved above this size-limit. For comet-size bodies (few km), though, the conclusion is the opposite. Most of the original comet-size planetesimals (if they existed - see Sect.~\ref{accretion} on the possible preferential formation of ``big'' objects in the streaming instability process) should have been destroyed. The survivors should have suffered multiple re-shaping collisions (Jutzi et al., 2017). This suggests that the comets that we see today, or at least their morphological structures (for instance the bi-lobed shapes), are  not primordial. Bi-lobed comets should be either the product of sub-catastrophic collisions (Jutzi and Benz, 2017), or fragments from catastrophic collisions (Schwartz et al., 2018). 

The collisional evolution of KBOs during the massive disk phase was addressed in Morbidelli and Rickman (2015). They found, not surprisingly, that the collisional activity is very intense in this phase, given the large amount of bodies in the disk (about 1,000 times the current population in the hot Kuiper belt - see sect.~\ref{hot}). The orbital eccentricities and inclinations where smaller than those characterizing the current Kuiper belt, but there was nevertheless a significant excitation due to the distant perturbations from Neptune and internal stirring from the Pluto-mass objects (whose number is estimated to have been $\sim 1,000$ - see sect.~\ref{res}). Given that the orbital periods where shorter than those of the current Kuiper belt, Morbidelli and Rickman estimated collisional velocities in the range 0.25-1 km/s (for comparison, the current collisional velocity for bodies with $a\sim 42$~AU, $e\sim 0.1$ and $i=10^\circ$ is $\sim 1$~km/s). Thus, collisions where clearly erosive and disruptive if the projectiles were large enough. 

The key question is then how long the disk remained in this state before getting dispersed by the giant planet instability and the migration of Neptune. When the original Nice model was proposed, it was suggested that the giant planet instability occurred $\sim 4$ Gy ago (namely $\sim 550$~My after gas removal from the disk), so to explain the origin of the so-called late heavy bombardment (LHB) of the Moon (Gomes et al., 2005; Bottke et al., 2012). In fact, at the time the most common interpretation of the LHB was that it corresponded to a sudden increase in the flux of impactors of the Earth-Moon system (Tera et al., 1974; Ryder, 2000). Thus, there was the need for a dynamical mechanism destabilizing reservoirs of small bodies after hundreds of My of relative stability: a late giant planet instability seemed perfect for that. But the evidence for an impact spike got weaker with the assessment of older impact events (Norman and Nemchin, 2014) and a reconsideration of the effects of impact age resetting (Boehnke and Harrison, 2016). Finally it was shown that the LHB can be explained as the tail of a bombardment ongoing since the time of terrestrial planet formation and declining over time, if one relaxes the constraint on the total amount of mass accreted by the Moon that had been (possibly incorrectly) deduced from the concentration of highly siderophile elements in the lunar mantle (Morbidelli et al., 2018). Thus there is no need that the trans-Neptunian disk remained massive for 550 My. 

Nesvorn\'y et al. (2018)  provided an upper limit of $\sim 100$~My for the dispersal of the trans-Neptunian disk. Their considerations are based on the existence of a relatively wide binary object (650~km separation) among the largest Jupiter's Trojans: Patroclus-Menoetius. They first verified that this binary would have had a probability of $\sim 70$\% to survive through the typical close encounters with the giant planets that transported objects from the trans-Neptunian disk to the Jupiter's coorbital region. This supports the idea that this binary, like all other Trojans, was captured from the trans-Neptunian disk. Then, they measured the rate at which similar binaries would have been dissociated by collisions in such disk. They concluded that the probability that Patroclus-Menoetius survived in the disk drops below 10\% if the disk lasted more than 100~My and below $7\times 10^{-5}$ if the disk lasted more than 400~My. Thus, it is very unlikely that the giant planet instability and the dispersal of the trans-Neptunian disk occurred at the time of the LHB. 

Clement et al. (2018) suggested that the giant planet instability occurred very early, within 10~My from gas photoevaporation. In this case most of the solid mass would have been removed from Mars' feeding zone while the planet was growing, thus explaining why this planet remained small. 

At the opposite end of the spectrum, Marty et al. (2017) argued for a significantly later instability, possibly around 100~My. By analyzing the xenon isotope composition of comet 67P-CG, they found that the original xenon in the Earth's atmosphere (called U-Xe), for long-time of unknown origin, is just the mixture of cometary Xe with meteoritic Xe. There is however no cometary Xe in the Earth's mantle. Only the meteoritic Xe can be found. This suggests that the cometary bombardment, associated with the giant planet instability and dispersal of the trans-Neptunian disk, occurred towards the end of the Earth's formation, possibly even after the Moon forming event and the crystallization of the terrestrial magma ocean. As the Moon-forming event is dated at $\sim 40$--100~My, this result argues for a comparable lifetime of the trans-Neptunian disk. 
 
Thus, it is fair to say that the lifetime of the trans-Neptunian disk is unknown and therefore it is not possible to assess a priori how much collisional evolution was suffered by the KBO population prior to its implantation in the Kuiper belt. 

\begin{figure}[t!]
\centerline{\includegraphics[height=8cm]{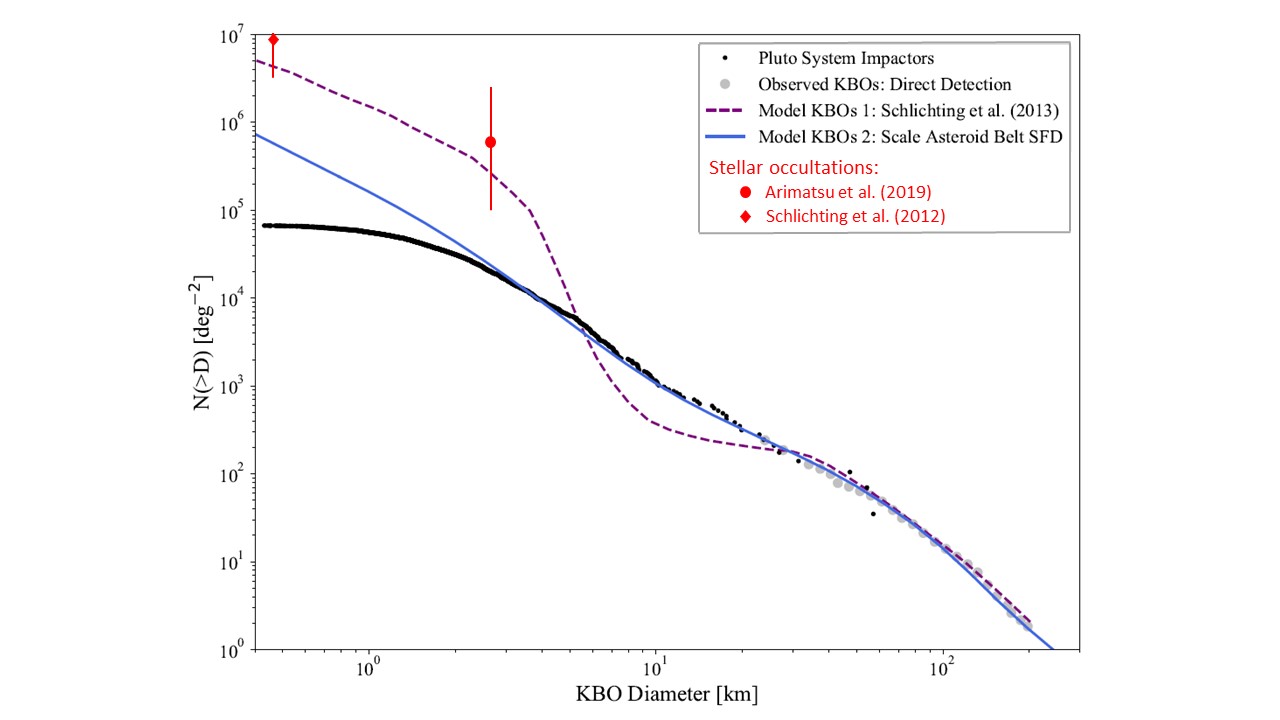}}
\caption{The black dots shows the size distribution of small KBOs determined by crater counting on Pluto and Charon (adapted from Singer et al., 2019). The gray dots show the size distribution of large KBOs determined by ground-based surveys. Both distributions have been translated into a density of bodies per sky unit area near the ecliptic. For comparison, the blue curve shows the size distribution of main belt asteroids, vertically rescaled. The dashed magenta curve shows the size distribution predicted in the coagulation model of Schlichting et al. (2013) (see sect.~\ref{accretion}). The red dots and vertical error bars show the sky densities of bodies larger than 500~m and 2.6~km in diameter as inferred by serendipitous stellar occultations (Schlichting et al., 2012; Arimatsu et al., 2019). Courtesy of A. Parker.}   
\label{Parker}
\end{figure}

However, thanks to the New Horizons mission, we may possibly turn the argument around. The mission allowed counting craters down to {a few km} in size on Pluto and Charon (Robbins et al., 2017; Singer et al., 2019). The observed crater size-frequency distribution can be turned into a projectile (i.e. KBO) size-frequency distribution using appropriate scaling laws (roughly relating the crater size to 10 times the projectile size). This deduced size-frequency distribution for small KBOs can then be smoothly branched with the size distribution of large KBOs determined by telescopic surveys from Earth. This is done in Fig.~\ref{Parker}. The figure compares the resulting size distribution of KBOs with the size distribution of main belt asteroids, which is well known down to sub-km sizes. Both are very similar {for $D\gtrsim 2$~km}.

{Below $D\sim 2$~km there are discordant results. Singer et al. (2019) claim that there is a paucity of craters on Pluto and Charon below $\sim 13$~km in diameter, which would imply a sharp turnover to a very shallow size distribution of Kuiper belt objects smaller than 1-2~km, as illustrated in Fig.~\ref{Parker}. Robbins et al. (2017), however, did not see this turnover, at least on the surface units imaged at the highest resolution (see Fig.~11 of Robbins et al. and compare it with Fig.~2b of Singer et al.). The reason for this discrepancy within the New Horizons team is not clear to us. The issue of the size distribution at km-scale becomes even more complicated with the result by Arimatsu et al. (2019) reporting the serendipitous detection by two telescopes of a stellar occultation by a trans-Neptunian object likely of 2.6~km in diameter at 30-50 AU. The statistical analysis of this detection implies a population of $D>2$~km bodies (red dot in Fig.~\ref{Parker}) at least an order of magnitude larger than that suggested by Pluto-Charon's crater record, but consistent with the population estimate deduced by another claimed stellar occultation detection by Schlichting et al. (2012) (red losange in Fig.~\ref{Parker}). }

{The issue is important and is related to both formation (see Sect.~2) and collisional evolution models. If there is really a deficit of sub-km objects as claimed by Singer et al., this means that very few small objects formed originally (consistent with streaming instability models) and that the size distribution of KBOs is not at collisional equilibrium (otherwise the small objects would have been produced in the collisional cascade, even if originally absent). If instead the km-sized bodies over-number the power-law extrapolation from 30-100~km, as claimed by Schlichting et al. and Arimatsu et al., this would suggest that planetesimals formed preferentially at these sizes as postulated in the collisional coagulation models. However, we have already commented in Sect. 2 that collisional coagulation models fail to reproduce the large-size end of the primordial trans-Neptunian disk.}

{Our tentative interpretation of these discordant observations is that the Kuiper belt populations have a normal, power-law size distribution similar to that of the asteroid belt. The crater record on Pluto and Charon is mostly dominated by bodies in the hot and cold components of the Kuiper belt (Greenstreet et al., 2015), while the occultation technique is mostly sensitive to scattered disk objects. There may be an order of magnitude ratio between the populations in the scattered disk and in the hot/cold Kuiper belt, thus explaining the apparent discordance in the detected population densities.  Notice that than 90\% of the Scattered Disk population at any given time is within 100 AU (Duncan and Levison 1997), so this is consistent with a claimed detection distance possibly of 50 AU (Arimatsu et al., 2019). Of course, this tentative interpretation needs to be confirmed through quantitative tests using an occultation survey simulator. At the time of writing, these tests have not yet been done. }

{Because of all these uncertainties, let's base our consideration on the size distribution of objects between 10 and 100~km, that is now quite safely assessed thanks to the New Horizons observations and telescopic surveys (see Fig.~\ref{Parker}). In this range, the} size distribution of the main asteroid belt for $D\lesssim 100$~km  is known to be the result of collisional equilibrium (Bottke et al., 2005). Thus, the similarity with the KBO size distribution suggests that the KBO population is also at collisional equilibrium below this size-limit. Nesvorn\'y \& Vokrouhlick\'y (2019) adopted an initial size distribution from the streaming instability simulations (Simon et al. 2016), and modelled the collisional evolution of the massive planetesimal disk at 20-30 AU (Figure \ref{si}a). Their target was to obtain the size distribution of dynamically hot KBOs and Jupiter Trojans (see Section 4.1.; Fraser et al., 2014)  They found that it is indeed possible that the initially rounded size distribution from the streaming instability collisionally evolved to match the distribution of hot KBOs and Jupiter's Trojans. There is a trade-off between the timescale of this evolution and the assumed strength of KBOs, with stronger (weaker) disruptions laws requiring longer (shorter) timescale. 
For instance, the result shown in Fig.~\ref{si}a has been obtained assuming a {specific energy of disruption for} KBOs equal to that computd by Benz and Asphaug (1999) for strong ice, divided by three. If the {the specific energy for disruption} is reduced by a factor of 10, the time required to acquire the observed size distribution decreases to 10~My.  
Thus, the disk lifetime thus cannot be inferred precisely from these simulations alone. 

Nesvorn\'y \& Vokrouhlick\'y (2019) also evaluated the effect of collisions on equal-size binaries and found that the survival depends on the binary size and separation (Figure \ref{si}b). For size/separation of equal-size binaries found in the cold Kuiper belt, the survival probability in the original, massive portion of the trans-Neptunian disk is 1-10\%. Remembering that this part of the disk is the progenitor of the hot population (sect.~\ref{hot}), this explains why the equal-size binary fraction among hot KBOs is relatively small. Moreover, the existence of a large fraction of binaries in the cold population, reinforces the idea that the disk beyond 40 AU was never massive (see sects.~\ref{accretion} and~\ref{cold}).          

\begin{figure}
\epsscale{0.45}
\plotone{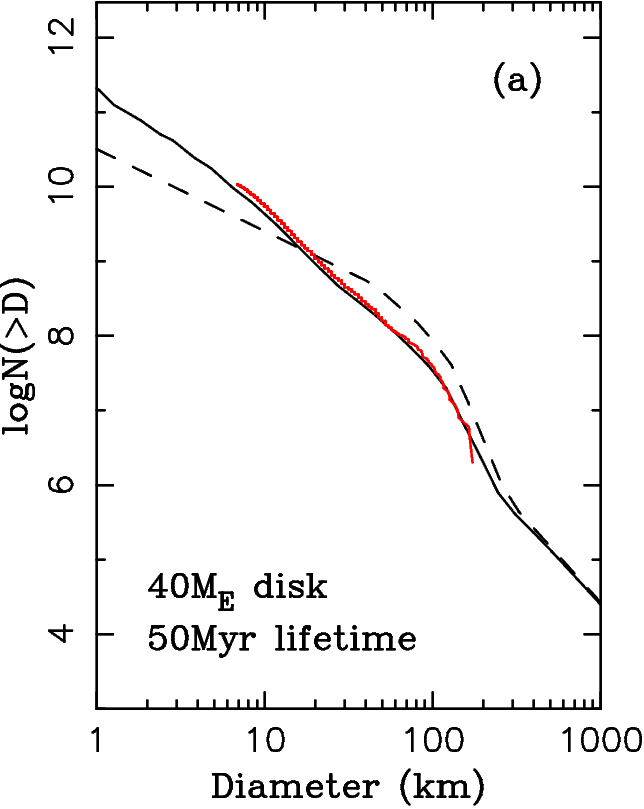}\hspace*{2.mm}
\plotone{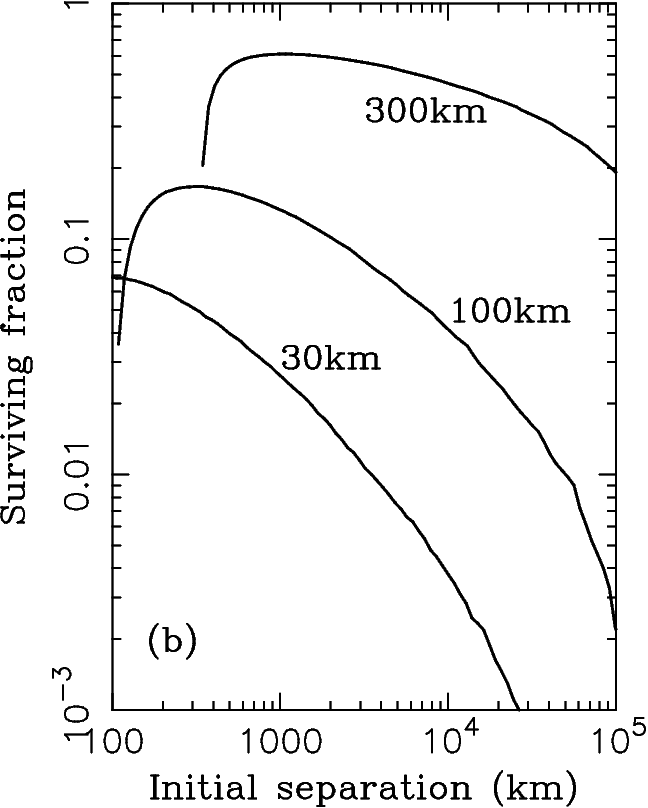}
\caption{Panel (a): collisional evolution of the outer planetesimal disk at 20-30 au. The dashed line shows the 
initial size distribution that was adopted from streaming instability simulations (Simon et al. 2016). It corresponds 
to the initial disk mass $M_{\rm disk}=40$~Earth masses. The solid line shows the size distribution after $t_{\rm disk}=50$ 
My of collisional grinding, when $M_{\rm disk}= 20$ Earth masses. The red line is the size distribution of known Jupiter Trojans 
(incomplete for $D<10$ km) scaled up by their implantation efficiency (Nesvorn\'y et al. 2013). Panel (b): collisional 
survival of equal-size binaries in the disk. The lines show the results for $t_{\rm disk}=50$~Myr and $R_1+R_2=30$, 100 
and 300 km, where $R_1$ and $R_2$ are the effective radii of two binary components. Figure from Nesvorn\'y and 
Vokrouhlick\'y (2019).}
\label{si}
\end{figure}

This result suggests that (i) the trans-Neptunian disk lasted $\sim 50$~My, which is consistent with the upper bound of $\sim 100$~My of Nesvorn\'y et al. (2018) and the indication of a post-Earth-formation cometary bombardment of Marty et al. (2017) and (ii) that the KBO population below 100~km in diameter has been strongly affected by collisions. In particular, the vast majority of cometary-size objects in the scenario would be fragments of larger objects, in a collisional cascade scenario.  

This result raises the question of whether catastrophic fragmentation could be consistent with the high porosity and low-temperature chemistry observed for 67P-CG and other comets.  Schwartz et al. (2018) showed that only a tiny fraction of the comet's material would have been heated by more than few degrees by the catastrophic collision that generated it.  A larger fraction of the parent body might have been substantially heated, but the heated material typically is not incorporated in macroscopic fragments. An analogous result was found for the porosity. It should be noted, however, that the collisions in Schwartz et al. (2018) are just above the catastrophic limit and the parent bodies considered are quite small: about 7 km in diameter.  Catastrophic collisions of parent bodies of tens of km in size have not yet been simulated, so we don't know whether the same results concerning heating and compaction apply.  Simulations of this kind can provide constraints on how big the progenitors of the collisional cascade leading to comet-sized objects could be. 

\section{Conclusions}

This review chapter has been divided in three parts. The initial part concerns the formation of KBOs; the central part concerns the origin of the different sub-components of the Kuiper belt, as well as the relationships with other population of objects, now distant from the Kuiper belt (e.g. Trojans, irregular satellites etc.). The final part discusses the collisional evolution of the KBOs coupled to their dynamical evolution.

The central part is the most consolidated one. Unlike the situation in 2008, now all the components of the Kuiper belt and the related distant populations are reproduced on quantitative grounds using a single model. In this model there are five giant planets on initially resonant and compact orbits, surrounded by a massive disk of planetesimals extended up to $\sim 30$~AU, with a low-mass extension to $\sim 45$~AU. Neptune migrates through the disk, dispersing its inner part and eventually triggering an instability in the planets' dynamics. At the instability one planet is ejected, all planets have encounters with at least another planet, which leads to orbital jumps. The final interaction of the planets with the dispersed disk stabilizes the planetary system and brings the planets to their current orbits.   

The power of this model has been to reproduce the observations in great details. Observers often ask for theoretical predictions, in additions to explanations of what is already known. This model implies several predictions, some of which already confirmed. It predicted that the size distributions of Jupiter's Trojans and of the hot Kuiper belt should be the same (Morbidelli et al., 2009). This has been confirmed in Fraser et al. (2014). From the expected no systematic differences between Jupiter-family and Oort cloud comets, it predicted the existence of JFCs with large D/H ratio, which has been confirmed by the ESA/Rosetta mission (Altwegg et al., 2015). It predicts that, with the exception of the cold population and the Sednoids, all other sub-populations of the Kuiper belt, the Trojans, the irregular satellites and the Oort cloud are parented; this can be checked by looking at spectral/color distributions, once irradiation and alteration effects are properly understood {(prediction already validated for the Jovian Trojan - irregular satellites connection by Graykowski and Jewitt, 2018)}. It predicts a deficit of small bodies in the Sednoid population compared to all other trans-Neptunian populations; this prediction can be tested with serendipitous stellar occultation experiments. Finally, it predicts that future surveys such as the LSST will not find anything really new in the Kuiper belt structure, but will just confirm and consolidate what is already known and explained. Only beyond 50~AU, where our current knowledge of the real population is sketchy, the LSST will unveil possible new features, but the model makes clear predictions of what will be found there (Nesvorn\'y et al., 2016).   

Given the current understanding of the orbital structure of the Kuiper belt, other expected contributions of LSST will be:
\begin{itemize}
\item { improving the characterization of the ``kernel'' of the cold Kuiper belt. This will help understanding if this structure is related to the sudden jump of Neptune's orbit, implying that the giant planet instability happened once Neptune had already reached $\sim$~28 AU (see sect.~\ref{cold})}.
\item {discovering/characterizing more Neptune's Trojans. The color distribution of Trojans, compared to the color distribution of the hot Kuiper belt can be diagnostic of the original color distribution in the trans-Neptunian disk and of the location of Neptune at the time of its orbital jump (see section~\ref{trojans})} 
\item understanding whether there is really a distant giant planet in the Solar System. In addition to potentially discovering the planet itself, the LSST will find a number of new fossilized scattered disk objects. With enough statistics it will become clear if there is an anomaly in orbital orientations that requires the existence of a distant planet to be explained.
\item constraining the original stellar cluster in which the Sun formed. By finding more Sednoids and characterizing their distribution in perihelion distance vs. semimajor axis, it will become possible to constrain quite precisely the density of the natal stellar cluster of the Solar System and even its lifetime. 
\end{itemize}

The origin of KBOs is not yet as clear as the orbital sculpting of the Kuiper belt. The realization that the original planetesimal population in the trans-Neptunian disk was about 1,000 times more populated than the current Kuiper belt defies the predictions of collisional coagulation models.  The streaming instability model, possibly combined with subsequent pebble accretion, seems to be more successful. The contrast between the massive disk within 30~AU and a low-mass disk in the 40-45~AU region also seems to suggest that large planetesimals formed in an instability process, which could act sporadically or frequently depending on local conditions. Nevertheless, the streaming instability model has to be improved on quantitative grounds to allow a pertinent comparison with the observations. Similarly, the fact that most objects of the cold population are binaries points to the gravitational collapse of rotating clouds of pebbles, as expected in the streaming instability model. The statistics of prograde/retrograde fractions of the equal-size binaries also support planetesimal formation by streaming instability. {The streaming instability model predicts that tight, equal size binaries (similar to Patroclus-Menoetius), not  easily dissociated during planetary encounters,  should be found in the hot population when observations reach the needed resolution. Verifying this prediction would also serve as a further confirmation that Jupiter Trojans have been captured from the original trans-Neptunian disk.}

The subsequent collisional evolution of KBOs depend crucially on the duration of the trans-Neptunian disk before its dynamical dispersal by the giant planet instability. This time could range from $\sim 0$ to $\sim 100$~My, with profound implications on how collisionally evolved the KBO population is. It will take time to clarify this issue. On one hand insight on the timing of the giant planet instability can be obtained by constraining the timing of the cometary bombardment of the Earth, given that some distinct cometary isotope signatures are now known. The paper by Marty et al. (2017) clearly establishes that a cometary bombardment existed on Earth, but the interpretation that the absence of cometary xenon in the Earth interior implies a post-Earth-formation bombardment has to be substantiated with geochemical models. On the other hand, achieving a better understanding of the size distribution produced in the streaming instability and comparing it with the observed size distribution can lead to an assessment of the importance of the collisional evolution. {The images of Ultima Thule taken during the flyby the New Horizons mission should eventually extend the size range on which the KBO size distribution is confidently determined from crater counting, thus solving the current controversy between Robbins et al. (2017), Singer et al. (2019) and Arimatsu et al. (2019). This would help constraining the collisional evolution of the Kuiper belt population. Unfortunately, at the time of writing of this review, the required high resolution images are not yet available.} 

In summary, although the Kuiper belt is no longer {\it Terra Incognita}, there is still room for new exploration and discoveries. Like the orbital structure of the Kuiper belt was determinant to constrain quite precisely the past orbital dynamics of the planets, its size distribution will provide crucial information to theorists to constrain formation models and to understand how collisional evolved the KBO population is. 

\acknowledgments 

\section{References}

\begin{itemize}

\item[--] A'Hearn, M.~F., Millis, R.~C., Schleicher, D.~O., Osip, D.~J., Birch, P.~V.\ 1995.\ The ensemble properties of comets: Results from narrowband photometry of 85 comets, 1976-1992..\ Icarus 118, 223-270. 

\item[--] Altwegg, K., and 31 colleagues 2015.\ 67P/Churyumov-Gerasimenko, a Jupiter family comet with a high D/H ratio.\ Science 347, 1261952. 

\item[--] Arimatsu, K., Tsumura, K., Usui, F., Shinnaka, Y., Ichikawa, K., Ootsubo, T., Kotani, T., Wada, T., Nagase, K., Watanabe, J.\ 2019.\ A kilometre-sized Kuiper belt object discovered by stellar occultation using amateur telescopes.\ Nature Astronomy, doi:10.1038/s41550-018-0685-8

\item[--] Batygin, K., Brown, M.~E., Fraser, W.~C.\ 2011.\ Retention of a Primordial Cold Classical Kuiper Belt in an Instability-Driven Model of Solar System Formation.\ The Astrophysical Journal 738, 13.

\item[--] Batygin, K., Brown, M.~E., Betts, H.\ 2012.\ Instability-driven Dynamical Evolution Model of a Primordially Five-planet Outer Solar System.\ The Astrophysical Journal 744, L3. 

\item[--] Batygin, K., Brown, M.~E.\ 2016.\ Evidence for a Distant Giant Planet in the Solar System.\ The Astronomical Journal 151, 22. 

\item[--] Benecchi, S.~D., Noll, K.~S., Grundy, W.~M., Buie, M.~W., Stephens, D.~C., Levison, H.~F.\ 2009.\ The correlated colors of transneptunian binaries.\ Icarus 200, 292-303.

\item[--] Benecchi, S.~D., Porter, S., Buie, M., Zangari, A., Verbiscer, A., Noll, K., Stern, S.~A., Spencer, J., Parker, A. 2019. The HST Lightcurve of (486958) 2014 MU69. Icarus, in press.

\item[--] Benz, W., Asphaug, E.\ 1999.\ Catastrophic Disruptions Revisited.\ Icarus 142, 5-20. 

\item[--] Birnstiel, T., Fang, M., Johansen, A.\ 2016.\ Dust Evolution and the Formation of Planetesimals.\ Space Science Reviews 205, 41-75. 
  
\item[--] Biver, N., Rauer, H., Despois, D., Moreno, R., Paubert, G., Bockel{\'e}e-Morvan, D., Colom, P., Crovisier, J., G{\'e}rard, E., Jorda, L.\ 1996.\ Substantial outgassing of CO from comet Hale-Bopp at large heliocentric distance.\ Nature 380, 137-139. 

\item[--]  Blum, J., and 21 colleagues 2017.\ Evidence for the formation of comet 67P/Churyumov-Gerasimenko through gravitational collapse of a bound clump of pebbles.\ Monthly Notices of the Royal Astronomical Society 469, S755-S773. 

\item[--] Boehnke, P., Harrison, T.M. 2016. Illusory Late Heavy Bombardment. Proceedings of the National Academy of Science 113, 10802-10806. 

\item[--] Bottke, W.~F., Durda, D.~D., Nesvorn{\'y}, D., Jedicke, R., Morbidelli, A., Vokrouhlick{\'y}, D., Levison, H.\ 2005.\ The fossilized size distribution of the main asteroid belt.\ Icarus 175, 111-140. 
  
\item[--] Bottke, W.~F., Nesvorn{\'y}, D., Vokrouhlick{\'y}, D., Morbidelli, A.\ 2010.\ The Irregular Satellites: The Most Collisionally Evolved Populations in the Solar System.\ The Astronomical Journal 139, 994-1014. 

\item[--] Bottke, W.F., Vokrouhlick\'y, D., Minton, D., Nesvorn\'y, D., Morbidelli, A., Brasser, R., Simonson, B., Levison, H.F. 2012. An Archaean heavy bombardment from a destabilized extension of the asteroid belt. Nature 485, 78-81.

\item[--] Brasser, R., Duncan, M.~J., Levison, H.~F.\ 2006.\ Embedded star clusters and the formation of the Oort Cloud.\ Icarus 184, 59-82. 

\item[--] Brasser, R., Duncan, M.~J., Levison, H.~F.\ 2007.\ Embedded star clusters and the formation of the Oort cloud. II. The effect of the primordial solar nebula.\ Icarus 191, 413-433. 

\item[--] Brasser, R., Duncan, M.~J., Levison, H.~F.\ 2008.\ Embedded star clusters and the formation of the Oort cloud. III. Evolution of the inner cloud during the Galactic phase.\ Icarus 196, 274-284. 

\item[--] Brasser, R., Morbidelli, A., Gomes, R., Tsiganis, K., Levison, H.~F.\ 2009.\ Constructing the secular architecture of the solar system II: the terrestrial planets.\ Astronomy and Astrophysics 507, 1053-1065. 

\item[--] Brasser, R., Duncan, M.~J., Levison, H.~F., Schwamb, M.~E., Brown, M.~E.\ 2012.\ Reassessing the formation of the inner Oort cloud in an embedded star cluster.\ Icarus 217, 1-19. 

\item[--] Brasser, R., Morbidelli, A.\ 2013.\ Oort cloud and Scattered Disc formation during a late dynamical instability in the Solar System.\ Icarus 225, 40-49. 

\item[--] Brauer, F., Dullemond, C.~P., Henning, T.\ 2008.\ Coagulation, fragmentation and radial motion of solid particles in protoplanetary disks.\ Astronomy and Astrophysics 480, 859-877. 
  
\item[--] Brown, M.~E.\ 2013.\ The Density of Mid-sized Kuiper Belt Object 2002 UX25 and the Formation of the Dwarf Planets.\ The Astrophysical Journal 778, L34. 

\item[--] Carrera, D., Gorti, U., Johansen, A., Davies, M.~B.\ 2017.\ Planetesimal Formation by the Streaming Instability in a Photoevaporating Disk.\ The Astrophysical Journal 839, 16. 

\item[--] Clement, M.~S., Kaib, N.~A., Raymond, S.~N., Walsh, K.~J.\ 2018.\ Mars' growth stunted by an early giant planet instability.\ Icarus 311, 340-356. 

\item[--] Davidsson, B.~J.~R., and 47 colleagues 2016.\ The primordial nucleus of comet 67P/Churyumov-Gerasimenko.\ Astronomy and Astrophysics 592, A63. 

\item[--] DeMeo, F.~E., Carry, B.\ 2014.\ Solar System evolution from compositional mapping of the asteroid belt.\ Nature 505, 629-634. 

\item[--] Dones, L., Weissman, P.~R., Levison, H.~F., Duncan, M.~J.\ 2004.\ Oort cloud formation and dynamics.\ Comets II 153-174. 

\item[--] Dr{\c a}{\.z}kowska, J., Alibert, Y., Moore, B.\ 2016.\ Close-in planetesimal formation by pile-up of drifting pebbles.\ Astronomy and Astrophysics 594, A105. 

\item[--] Dr{\c a}{\.z}kowska, J., Alibert, Y.\ 2017.\ Planetesimal formation starts at the snow line.\ Astronomy and Astrophysics 608, A92. 

\item[--] Duncan, M., Quinn, T., Tremaine, S.\ 1987.\ The formation and extent of the solar system comet cloud.\ The Astronomical Journal 94, 1330-1338. 

\item[--] Duncan, M.~J., Levison, H.~F.\ 1997.\ A scattered comet disk and the origin of Jupiter family comets.\ Science 276, 1670-1672. 

\item[--] Fernandez, J.~A., Ip, W.-H.\ 1984.\ Some dynamical aspects of the accretion of Uranus and Neptune - The exchange of orbital angular momentum with planetesimals.\ Icarus 58, 109-120. 

\item[--] Fraser, W.~C., Brown, M.~E., Schwamb, M.~E.\ 2010.\ The luminosity function of the hot and cold Kuiper belt populations.\ Icarus 210, 944-955.

\item[--] Fraser, W.~C., Brown, M.~E., Morbidelli, A., Parker, A., Batygin, K.\ 2014.\ The Absolute Magnitude Distribution of Kuiper Belt Objects.\ The Astrophysical Journal 782, 100. 

\item[--] Fraser, W.~C., and 21 colleagues 2017.\ All planetesimals born near the Kuiper belt formed as binaries.\ Nature Astronomy 1, 0088. 

\item[--] Gladman, B., Marsden, B.~G., Vanlaerhoven, C.\ 2008.\ Nomenclature in the Outer Solar System.\ The Solar System Beyond Neptune 43-57. 

\item[--] Gladman, B., and 10 colleagues 2012.\ The Resonant Trans-Neptunian Populations.\ The Astronomical Journal 144, 23. 

\item[--] Goldreich, P., Ward, W.~R.\ 1973.\ The Formation of Planetesimals.\ The Astrophysical Journal 183, 1051-1062. 

\item[--] Goldreich, P., Lithwick, Y., Sari, R.\ 2002.\ Formation of Kuiper-belt binaries by dynamical friction and three-body encounters.\ Nature 420, 643-646. 

\item[--] Gomes, R.~S.\ 2003.\ The origin of the Kuiper Belt high-inclination population.\ Icarus 161, 404-418. 

\item[--] Gomes, R., Levison, H.~F., Tsiganis, K., Morbidelli, A.\ 2005.\ Origin of the cataclysmic Late Heavy Bombardment period of the terrestrial planets.\ Nature 435, 466-469. 

\item[--] Gomes, R.~S., Gallardo, T., Fern{\'a}ndez, J.~A., Brunini, A.\ 2005b.\ On The Origin of The High-Perihelion Scattered Disk: The Role of The Kozai Mechanism And Mean Motion Resonances.\ Celestial Mechanics and Dynamical Astronomy 91, 109-129. 

\item[--] Gomes, R.~S.\ 2011.\ The origin of TNO 2004 XR $_{190}$ as a primordial scattered object.\ Icarus 215, 661-668. 

\item[--] Gomes, R., Nesvorn{\'y}, D.\ 2016.\ Neptune trojan formation during planetary instability and migration.\ Astronomy and Astrophysics 592, A146. 

\item[--] Graykowski, A., Jewitt, D.\ 2018.\ Colors and Shapes of the Irregular Planetary Satellites.\ The Astronomical Journal 155, 184. 
  
\item[--] Greenstreet, S., Gladman, B., McKinnon, W.~B.\ 2015.\ Impact and cratering rates onto Pluto.\ Icarus 258, 267-288. 

\item[--] Grundy, W. M. et al. (10 coauthors) 2019, Mutual Orbit Orientations of Transneptunian Binaries, Icarus, submitted

\item[--] G{\"u}ttler, C., Blum, J., Zsom, A., Ormel, C.~W., Dullemond, C.~P.\ 2009.\ The first phase of protoplanetary dust growth: The bouncing barrier.\ Geochimica et Cosmochimica Acta Supplement 73, A482. 

\item[--] Hahn, J.~M., Malhotra, R.\ 2005.\ Neptune's Migration into a Stirred-Up Kuiper Belt: A Detailed Comparison of Simulations to Observations.\ The Astronomical Journal 130, 2392-2414.  

\item[--] Hainaut, O.~R., Boehnhardt, H., Protopapa, S.\ 2012.\ Colours of minor bodies in the outer solar system. II. A statistical analysis revisited.\ Astronomy and Astrophysics 546, A115. 
  
\item[--] Ida, S., Guillot, T.\ 2016.\ Formation of dust-rich planetesimals from sublimated pebbles inside of the snow line.\ Astronomy and Astrophysics 596, L3. 

\item[--] Jacquet, E., Balbus, S., Latter, H. 2011. On linear dust-gas streaming instabilities in protoplanetary discs. Monthly Notices of the Royal Astronomical Society 415, 3591-3598. 

\item[--] Jehin, E., Manfroid, J., Hutsem{\'e}kers, D., Cochran, A., Zucconi, J.-M., Schulz, R., Arpigny, C.\ 2008.\ Carbon and Nitrogen Isotopic Ratios in Comets.\ Asteroids, Comets, Meteors 2008 1405, 8339.

\item[--] Jewitt, D., Sheppard, S.\ 2005.\ Irregular Satellites in the Context of Planet Formation.\ Space Science Reviews 116, 441-455. 

\item[--] Jewitt, D.\ 2018.\ The Trojan Color Conundrum.\ The Astronomical Journal 155, 56. 

\item[--] J{\'{\i}}lkov{\'a}, L., Portegies Zwart, S., Pijloo, T., Hammer, M.\ 2015.\ How Sedna and family were captured in a close encounter with a solar sibling.\ Monthly Notices of the Royal Astronomical Society 453, 3157-3162. 

\item[--] Johansen, A., Oishi, J.~S., Mac Low, M.-M., Klahr, H., Henning, T., Youdin, A.\ 2007.\ Rapid planetesimal formation in turbulent circumstellar disks.\ Nature 448, 1022-1025. 

\item[--] Johansen, A., Youdin, A., Mac Low, M.-M.\ 2009.\ Particle Clumping and Planetesimal Formation Depend Strongly on Metallicity.\ The Astrophysical Journal 704, L75-L79. 

\item[--] Johansen, A., Mac Low, M.M., Lacerda, P., Bizzarro, M. 2015. Growth of asteroids, planetary embryos, and Kuiper belt objects by chondrule accretion. Science Advances 1, 1500109. 

\item[--] Jutzi, M., Benz, W.\ 2017.\ Formation of bi-lobed shapes by sub-catastrophic collisions. A late origin of comet 67P's structure.\ Astronomy and Astrophysics 597, A62. 

\item[--] Jutzi, M., Benz, W., Toliou, A., Morbidelli, A., Brasser, R.\ 2017.\ How primordial is the structure of comet 67P?. Combined collisional and dynamical models suggest a late formation.\ Astronomy and Astrophysics 597, A61. 

\item[--] Kenyon, S.~J., Bromley, B.~C.\ 2004.\ The Size Distribution of Kuiper Belt Objects.\ The Astronomical Journal 128, 1916-1926. 

\item[--] Kenyon, S.~J., Bromley, B.~C.\ 2004b.\ Stellar encounters as the origin of distant Solar System objects in highly eccentric orbits.\ Nature 432, 598-602. 

\item[--] Kenyon, S.~J., Bromley, B.~C., O'Brien, D.~P., Davis, D.~R.\ 2008.\ Formation and Collisional Evolution of Kuiper Belt Objects.\ The Solar System Beyond Neptune 293-313. 

\item[--] Kenyon, S.~J., Bromley, B.~C.\ 2012.\ Coagulation Calculations of Icy Planet Formation at 15-150 AU: A Correlation between the Maximum Radius and the Slope of the Size Distribution for Trans-Neptunian Objects.\ The Astronomical Journal 143, 63. 

\item[--] Kleine, T., Mezger, K., Palme, H., Scherer, E., M{\"u}nker, C.\ 2005.\ Early core formation in asteroids and late accretion of chondrite parent bodies: Evidence from $^{182}$Hf- $^{182}$W in CAIs, metal-rich chondrites, and iron meteorites.\ Geochimica et Cosmochimica Acta 69, 5805-5818.

\item[--] Kruijer, T.~S., Touboul, M., Fischer-G{\"o}dde, M., Bermingham, K.~R., Walker, R.~J., Kleine, T.\ 2014.\ Protracted core formation and rapid accretion of protoplanets.\ Science 344, 1150-1154.  

\item[--] Lambrechts, M., Johansen, A.\ 2012.\ Rapid growth of gas-giant cores by pebble accretion.\ Astronomy and Astrophysics 544, A32. 

\item[--] Lambrechts, M., Johansen, A.\ 2014.\ Forming the cores of giant planets from the radial pebble flux in protoplanetary discs.\ Astronomy and Astrophysics 572, A107. 

\item[--] Lambrechts, M., Morbidelli, A.\ 2016.\ Reconstructing the size distribution of the small body population in the Solar System.\ AAS/Division for Planetary Sciences Meeting Abstracts \#48 48, 105.08. 

\item[--] Lawler, S.~M., Kavelaars, J.~J., Alexandersen, M., Bannister, M.~T., Gladman, B., Petit, J.-M., Shankman, C.\ 2018a.\ OSSOS: X. How to use a Survey Simulator: Statistical Testing of Dynamical Models Against the Real Kuiper Belt.\ Frontiers in Astronomy and Space Sciences 5, 14. 

\item[--] Lawler, S.~M., and 10 colleagues 2018b.\ OSSOS: XIII. Fossilized Resonant Dropouts Imply Neptune's Migration was Grainy and Slow.\ ArXiv e-prints arXiv:1808.02618. 

\item[--] Levison, H.~F., Morbidelli, A.\ 2003.\ The formation of the Kuiper belt by the outward transport of bodies during Neptune's migration.\ Nature 426, 419-421. 

\item[--] Levison, H.~F., Morbidelli, A., Van Laerhoven, C., Gomes, R., Tsiganis, K.\ 2008.\ Origin of the structure of the Kuiper belt during a dynamical instability in the orbits of Uranus and Neptune.\ Icarus 196, 258-273. 

\item[--] Levison, H.~F., Bottke, W.~F., Gounelle, M., Morbidelli, A., Nesvorn{\'y}, D., Tsiganis, K.\ 2009.\ Contamination of the asteroid belt by primordial trans-Neptunian objects.\ Nature 460, 364-366. 

\item[--] Levison, H.~F., Morbidelli, A., Tsiganis, K., Nesvorn{\'y}, D., Gomes, R.\ 2011.\ Late Orbital Instabilities in the Outer Planets Induced by Interaction with a Self-gravitating Planetesimal Disk.\ The Astronomical Journal 142, 152. 

\item[--] Malhotra, R.\ 1993.\ The origin of Pluto's peculiar orbit.\ Nature 365, 819-821. 

\item[--] Malhotra, R.\ 1995.\ The Origin of Pluto's Orbit: Implications for the Solar System Beyond Neptune.\ The Astronomical Journal 110, 420. 

\item[--] Marty, B., and 29 colleagues 2017.\ Xenon isotopes in 67P/Churyumov-Gerasimenko show that comets contributed to Earth's atmosphere.\ Science 356, 1069-1072. 

\item[--] Marzari, F., Scholl, H., Murray, C., Lagerkvist, C.\ 2002.\ Origin and Evolution of Trojan Asteroids.\ Asteroids III 725-738. 

\item[--] Morbidelli, A., Levison, H.~F.\ 2004.\ Scenarios for the Origin of the Orbits of the Trans-Neptunian Objects 2000 CR$_{105}$ and 2003 VB$_{12}$ (Sedna).\ The Astronomical Journal 128, 2564-2576. 

\item[--] Morbidelli, A., Levison, H.~F., Tsiganis, K., Gomes, R.\ 2005.\ Chaotic capture of Jupiter's Trojan asteroids in the early Solar System.\ Nature 435, 462-465. 

\item[--] Morbidelli, A., Tsiganis, K., Crida, A., Levison, H.~F., Gomes, R.\ 2007.\ Dynamics of the Giant Planets of the Solar System in the Gaseous Protoplanetary Disk and Their Relationship to the Current Orbital Architecture.\ The Astronomical Journal 134, 1790-1798. 

\item[--] Morbidelli, A., Levison, H.~F., Gomes, R.\ 2008.\ The Dynamical Structure of the Kuiper Belt and Its Primordial Origin.\ The Solar System Beyond Neptune 275-292. 

\item[--] Morbidelli, A., Levison, H.~F., Bottke, W.~F., Dones, L., Nesvorn{\'y}, D.\ 2009.\ Considerations on the magnitude distributions of the Kuiper belt and of the Jupiter Trojans.\ Icarus 202, 310-315. 
  
\item[--] Morbidelli, A., Brasser, R., Gomes, R., Levison, H.~F., Tsiganis, K.\ 2010.\ Evidence from the Asteroid Belt for a Violent Past Evolution of Jupiter's Orbit.\ The Astronomical Journal 140, 1391-1401. 

\item[--] Morbidelli, A., Rickman, H.\ 2015.\ Comets as collisional fragments of a primordial planetesimal disk.\ Astronomy and Astrophysics 583, A43. 

\item[--] Morbidelli, A., Nesvorn\'y, D., Laurenz, V., Marchi, S., Rubie, D.~C., Elkins-Tanton, L., Wieczorek, M., Jacobson, S.\ 2018.\ The timeline of the lunar bombardment: Revisited.\ Icarus 305, 262-276. 

\item[--] Nesvorn{\'y}, D., Dones, L.\ 2002.\ How Long-Lived Are the Hypothetical Trojan Populations of Saturn, Uranus, and Neptune?.\ Icarus 160, 271-288. 

\item[--] Nesvorn{\'y}, D., Vokrouhlick{\'y}, D., Morbidelli, A.\ 2007.\ Capture of Irregular Satellites during Planetary Encounters.\ The Astronomical Journal 133, 1962-1976. 

\item[--] Nesvorn{\'y}, D., Youdin, A.~N., Richardson, D.~C.\ 2010.\ Formation of Kuiper Belt Binaries by Gravitational Collapse.\ The Astronomical Journal 140, 785-793. 

\item[--] Nesvorn{\'y}, D.\ 2011.\ Young Solar System's Fifth Giant Planet?.\ The Astrophysical Journal 742, L22. 

\item[--] Nesvorn{\'y}, D., Vokrouhlick{\'y}, D., Bottke, W.~F., Noll, K., Levison, H.~F.\ 2011.\ Observed Binary Fraction Sets Limits on the Extent of Collisional Grinding in the Kuiper Belt.\ The Astronomical Journal 141, 159. 

\item[--] Nesvorn{\'y}, D., Morbidelli, A.\ 2012.\ Statistical Study of the Early Solar System's Instability with Four, Five, and Six Giant Planets.\ The Astronomical Journal 144, 117. 

\item[--] Nesvorn{\'y}, D., Vokrouhlick{\'y}, D., Morbidelli, A.\ 2013.\ Capture of Trojans by Jumping Jupiter.\ The Astrophysical Journal 768, 45. 

\item[--] Nesvorn{\'y}, D., Vokrouhlick{\'y}, D., Deienno, R.\ 2014.\ Capture of Irregular Satellites at Jupiter.\ The Astrophysical Journal 784, 22. 

\item[--] Nesvorn{\'y}, D.\ 2015a.\ Jumping Neptune Can Explain the Kuiper Belt Kernel.\ The Astronomical Journal 150, 68. 

\item[--] Nesvorn{\'y}, D.\ 2015b.\ Evidence for Slow Migration of Neptune from the Inclination Distribution of Kuiper Belt Objects.\ The Astronomical Journal 150, 73. 

\item[--] Nesvorn{\'y}, D., Vokrouhlick{\'y}, D.\ 2016.\ Neptune's Orbital Migration Was Grainy, Not Smooth.\ The Astrophysical Journal 825, 94. 

\item[--] Nesvorn{\'y}, D., Vokrouhlick{\'y}, D., Roig, F.\ 2016.\ The Orbital Distribution of Trans-Neptunian Objects Beyond 50 au.\ The Astrophysical Journal 827, L35. 

\item[--] Nesvorn{\'y}, D., Vokrouhlick{\'y}, D., Dones, L., Levison, H.~F., Kaib, N., Morbidelli, A.\ 2017.\ Origin and Evolution of Short-period Comets.\ The Astrophysical Journal 845, 27. 

\item[--] Nesvorn{\'y}, D., Vokrouhlick{\'y}, D., Bottke, W.~F., Levison, H.~F.\ 2018.\ Evidence for very early migration of the Solar System planets from the Patroclus-Menoetius binary Jupiter Trojan.\ Nature Astronomy 2, 878-882. 

\item[--] Nesvorn{\'y}, D., Parker, J., Vokrouhlick{\'y}, D.\ 2018b.\ Bi-lobed Shape of Comet 67P from a Collapsed Binary.\ The Astronomical Journal 155, 246. 
  
\item[--] Nesvorn\'y, D., Li, R., Youdin, N. Y., Simon, J. B., Grundy, W. M., 2019. Trans-Neptunian Binaries Provide Evidence for Planetesimal Formation by the Streaming Instability, Nature Astronomy, submitted

\item[--] Nesvorn\'y, D., Vokrouhlick\'y, D. 2019. Binary Survival in the Outer Solar System, Icarus, submitted. 

\item[--] Noll, K.~S., Levison, H.~F., Grundy, W.~M., Stephens, D.~C.\ 2006.\ Discovery of a binary Centaur.\ Icarus 184, 611-618. 
  
\item[--] Noll, K.~S., Grundy, W.~M., Chiang, E.~I., Margot, J.-L., Kern, S.~D.\ 2008.\ Binaries in the Kuiper Belt.\ The Solar System Beyond Neptune 345-363. 

\item[--] Norman, M.D., Nemchin, A.A. 2014. A 4.2 billion year old impact basin on the Moon: U-Pb dating of zirconolite and apatite in lunar melt rock 67955. Earth and Planetary Science Letters 388, 387-398. 

\item[--] Oort, J.~H.\ 1950.\ The structure of the cloud of comets surrounding the Solar System and a hypothesis concerning its origin.\ Bulletin of the Astronomical Institutes of the Netherlands 11, 91-110. 

\item[--] Pan, M., Sari, R.\ 2005.\ Shaping the Kuiper belt size distribution by shattering large but strengthless bodies.\ Icarus 173, 342-348. 

\item[--] Parker, A.~H., Kavelaars, J.~J.\ 2010.\ Destruction of Binary Minor Planets During Neptune Scattering.\ The Astrophysical Journal 722, L204-L208. 

\item[--] Petit, J.-M., Mousis, O.\ 2004.\ KBO binaries: how numerous were they?.\ Icarus 168, 409-419. 

\item[--] Petit, J.-M., Kavelaars, J.~J., Gladman, B., Loredo, T.\ 2008.\ Size Distribution of Multikilometer Transneptunian Objects.\ The Solar System Beyond Neptune 71-87. 

\item[--] Petit, J.-M., and 16 colleagues 2011.\ The Canada-France Ecliptic Plane Survey--Full Data Release: The Orbital Structure of the Kuiper Belt.\ The Astronomical Journal 142, 131. 

\item[--] Prialnik, D., Sarid, G., Rosenberg, E.~D., Merk, R.\ 2009.\ Thermal and Chemical Evolution of Comet Nuclei and Kuiper Belt Objects.\ Origin and Early Evolution of Comet Nuclei 147. 

\item[--] Ribeiro de Sousa, R., Gomes, R., Morbidelli, A., Vieira Neto, E.\ 2018.\ Dynamical effects on the classical Kuiper Belt during the excited-Neptune model.\ ArXiv e-prints arXiv:1808.02146. 

\item[--] Robbins, S.~J., and 29 colleagues 2017.\ Craters of the Pluto-Charon system.\ Icarus 287, 187-206. 

\item[--] Ryder, G. 2002. Mass flux in the ancient Earth-Moon system and benign implications for the origin of life on Earth. Journal of Geophysical Research (Planets) 107, 5022-1.

\bibitem[Schlichting and Sari(2008)]{2008ApJ...686..741S} Schlichting, H.~E., Sari, R.\ 2008.\ The Ratio of Retrograde to Prograde Orbits: A Test for Kuiper Belt Binary Formation Theories.\ The Astrophysical Journal 686, 741-747. 

\item[--] Schlichting, H.~E., Sari, R.\ 2011.\ Runaway Growth During Planet Formation: Explaining the Size Distribution of Large Kuiper Belt Objects.\ The Astrophysical Journal 728, 68. 

\item[--] Schlichting, H.~E., Ofek, E.~O., Sari, R., Nelan, E.~P., Gal-Yam, A., Wenz, M., Muirhead, P., Javanfar, N., Livio, M.\ 2012.\ Measuring the Abundance of Sub-kilometer-sized Kuiper Belt Objects Using Stellar Occultations.\ The Astrophysical Journal 761, 150. 

\item[--] Schlichting, H.~E., Fuentes, C.~I., Trilling, D.~E.\ 2013.\ Initial Planetesimal Sizes and the Size Distribution of Small Kuiper Belt Objects.\ The Astronomical Journal 146, 36. 

\item[--] Schoonenberg, D., Ormel, C.~W.\ 2017.\ Planetesimal formation near the snowline: in or out?.\ Astronomy and Astrophysics 602, A21. 

\item[--] Schwartz, S.~R., Michel, P., Jutzi, M., Marchi, S., Zhang, Y., Richardson, D.~C.\ 2018.\ Catastrophic disruptions as the origin of bilobate comets.\ Nature Astronomy 2, 379-382.

\item[--] Shakura, N.~I., Sunyaev, R.~A.\ 1973.\ Black holes in binary systems. Observational appearance..\ Astronomy and Astrophysics 24, 337-355. 

\item[--] Shannon, A., Wu, Y., Lithwick, Y.\ 2016.\ Forming the Cold Classical Kuiper Belt in a Light Disk.\ The Astrophysical Journal 818, 175. 

\item[--] Sheppard, S.~S., Trujillo, C.~A.\ 2010.\ Detection of a Trailing (L5) Neptune Trojan.\ Science 329, 1304. 

\item[--] Sheppard, S.~S.\ 2012.\ The Color Differences of Kuiper Belt Objects in Resonance with Neptune.\ The Astronomical Journal 144, 169. 

\item[--] Shoemaker, E.~M., Wolfe, R.~F.\ 1984.\ Evolution of the Uranus-Neptune Planetesimal Swarm.\ Lunar and Planetary Science Conference 15, 780-781. 

\item[--] Simon, J.B., Armitage, P.J., Li, R., Youdin, A.N. 2016. The Mass and Size Distribution of Planetesimals Formed by the Streaming Instability. I. The Role of Self-gravity. The Astrophysical Journal 822, 55. 

\item[--] Singer, K.~N., and 22 colleagues 2019.\ Impact Craters on Pluto and Charon Indicate a Deficit of Small Kuiper Belt Objects. Science 363, 955.
  
\item[--] Sosa, A., Fern{\'a}ndez, J.~A.\ 2011.\ Masses of long-period comets derived from non-gravitational effects - analysis of the computed results and the consistency and reliability of the non-gravitational parameters.\ Monthly Notices of the Royal Astronomical Society 416, 767-782. 

\item[--] Stern, S.~A.\ 1991.\ On the number of planets in the outer solar system - Evidence of a substantial population of 1000-km bodies.\ Icarus 90, 271-281. 

\item[--] Tera, F., Papanastassiou, D.A., Wasserburg, G.J. 1974. Isotopic evidence for a terminal lunar cataclysm. Earth and Planetary Science Letters 22, 1-21.

\item[--] Thommes, E.~W., Duncan, M.~J., Levison, H.~F.\ 2002.\ The Formation of Uranus and Neptune among Jupiter and Saturn.\ The Astronomical Journal 123, 2862-2883. 

\item[--] Trujillo, C.~A., Jewitt, D.~C., Luu, J.~X.\ 2000.\ Population of the Scattered Kuiper Belt.\ The Astrophysical Journal 529, L103-L106. 

\item[--] Trujillo, C.~A., Brown, M.~E.\ 2002.\ A Correlation between Inclination and Color in the Classical Kuiper Belt.\ The Astrophysical Journal 566, L125-L128. 

\item[--] Trujillo, C.~A., Sheppard, S.~S.\ 2014.\ A Sedna-like body with a perihelion of 80 astronomical units.\ Nature 507, 471-474. 

\item[--] Tsiganis, K., Gomes, R., Morbidelli, A., Levison, H.~F.\ 2005.\ Origin of the orbital architecture of the giant planets of the Solar System.\ Nature 435, 459-461. 

\item[--] Villeneuve, J., Chaussidon, M., Libourel, G.\ 2009.\ Homogeneous Distribution of $^{26}$Al in the Solar System from the Mg Isotopic Composition of Chondrules.\ Science 325, 985.

\item[--] Vokrouhlick{\'y}, D., Bottke, W.~F., Nesvorn{\'y}, D.\ 2016.\ Capture of Trans-Neptunian Planetesimals in the Main Asteroid Belt.\ The Astronomical Journal 152, 39. 

\item[--] Volk, K., Malhotra, R.\ 2008.\ The Scattered Disk as the Source of the Jupiter Family Comets.\ The Astrophysical Journal 687, 714-725. 

\item[--] Weidenschilling, S.~J.\ 1977.\ Aerodynamics of solid bodies in the solar nebula.\ Monthly Notices of the Royal Astronomical Society 180, 57-70. 
  
\item[--] Weidenschilling, S.~J.\ 1995.\ Can Gravitational Instability Form Planetesimals?.\ Icarus 116, 433-435.

\item[--] Wong, I., Brown, M.~E.\ 2016.\ A Hypothesis for the Color Bimodality of Jupiter Trojans.\ The Astronomical Journal 152, 90. 

\item[--] Yang, C.-C., Johansen, A., Carrera, D.\ 2017.\ Concentrating small particles in protoplanetary disks through the streaming instability.\ Astronomy and Astrophysics 606, A80. 

\item[--] Youdin, A.~N., Shu, F.~H.\ 2002.\ Planetesimal Formation by Gravitational Instability.\ The Astrophysical Journal 580, 494-505. 

\item[--] Youdin, A.~N., Goodman, J.\ 2005.\ Streaming Instabilities in Protoplanetary Disks.\ The Astrophysical Journal 620, 459-469.

\end{itemize}
\end{document}